# CHARACTERISATION OF MARTIAN DUST AEROSOL PHASE FUNCTION FROM SKY RADIANCE MEASUREMENTS BY MSL ENGINEERING CAMERAS

H. Chen-Chen*, S. Pérez-Hoyos, A. Sánchez-Lavega

[a]Departamento de Física Aplicada I, Escuela de Ingeniería de Bilbao, Universidad del País Vasco (UPV/EHU). Bilbao 48013, Spain

* To whom correspondence should be addressed at: **hao.chen@ehu.eus**

## Abstract

Dust is the main driver of Mars' atmospheric variability. The determination of Martian dust aerosol properties is of high relevance for radiative modelling and calculating its weather forcing. In particular, the light scattering behaviour at intermediate and large scattering angles can provide valuable information regarding the airborne dust particle shape. The angular distribution of sky brightness observed by the Mars Science Laboratory engineering cameras (Navcam and Hazcam) is used here to characterise the atmospheric dust single scattering phase function and to constrain the shape of the particles. An iterative radiative transfer based retrieval method was implemented in order to determine the aerosol modelling parameters which best reproduce the observed sky radiance as a function of the scattering angle in the solar almucantar plane. The aerosol models considered in this study for retrieving dust radiative properties were an analytical three term Double Henyey-Greenstein (DHG) phase function, T-matrix calculations for cylindrical particles with different diameter-to-length (D/L) aspect ratios and experimental phase functions from laboratory measurements of several Martian dust analogue samples. Results of this study returned mean DHG phase function parameter values $g_1$ = 0.889±0.098, $g_2$ = 0.094±0.250, $\alpha$ = 0.743±0.106; generating a phase function with an asymmetry parameter of g = 0.673±0.081 (in line with Wolff et al., 2009). Although differences were observed during the low opacity aphelion season (lower forward scattering values, presence of a peak in the backward region) compared to the rest of the year, no clear evidences of seasonal behaviour or interannual variability were derived. The obtained average D/L aspect ratios for T-matrix calculated cylindrical particles were 0.70±0.20 and 1.90±0.20 (similar to Wolff et al., 2001), and the best fitting Martian dust analogue corresponded to the basalt sample (in agreement with Dabrowska et al., 2015).

## 1. Introduction

Dust is the main driver of the Martian atmospheric variability; it has a direct impact on the atmospheric thermal structure and provides forcing to its dynamics by absorbing and scattering solar radiation (Gierasch and Goody, 1972; Pollack et al., 1979). The importance of dust in Mars' climate has focused intensive research efforts on characterising its microphysical (e.g. particle size, shape, composition) and radiative (extinction efficiency, single scattering albedo, phase function) properties. Comprehensive reviews of results from previous studies can be found in Korablev et al. (2005), Smith et al. (2008). Because the radiative parameters depend on the microphysical properties of dust particles, the inversion of observation data to retrieve these parameters represents a challenging task (Pollack et al., 1995; Clancy et al., 2003; Wolff et al., 2009).

The single scattering phase function describes the angular distribution of the scattered light by aerosols and it is strongly influenced by the size and shape of the particles. In particular, the light scattering behaviour at intermediate and large scattering angles can provide relevant information on the aerosol particle shape (Kaufman et al., 1994). The characterisation of the particle shape is relevant as it affects the estimates of other parameters, such as the aerosol column optical thickness and the imaginary part of the refractive index (Dlugach et al., 2002). While light scattering calculations for spherical particles are straightforward by using the Lorenz-Mie theory (e.g. Hansen and Travis, 1974), calculations considering realistic dispersions of non-spherical particles may result very complex and computationally demanding (Dubovik et al., 2006; Yang et al., 2007; Yurkin and Hoekstra, 2007).

Retrievals of Martian atmospheric dust phase function and constraint of particle shape have been performed by several authors using both orbital observations and surface-based sky imaging data. Chýlek and Grams (1978) used a non-spherical randomly oriented particle model to fit Mariner 9 reflectance data during the 1971 Mars dust storm. Pollack et al. (1977, 1995) analysed Viking Lander sky images and used a semi-empirical theory to model scattering properties by non-spherical particles (Pollack and Cuzzi, 1980) for fitting the observations. They retrieved a modest peak in the backscattering region that suggested internal reflections by sharp corners within the particle's geometry associated to fluffy aggregates. Sky brightness data obtained by the Imager for Mars Pathfinder (IMP) were fitted with multiple scattering radiative transfer calculations to retrieve dust properties, including the single scattering phase function (Tomasko et al., 1999), and presented good agreements with plate-like particles (Markiewicz et al., 1999). Wolff et al. (2001) compared Mars Global Surveyor Thermal Emission Spectrometer (MGS TES) dust phase functions retrievals using radiative transfer simulations with T-matrix computations of non-spherical aerosols (Mishchenko et al., 1998) and obtained best-fits for randomly oriented cylinders with diameter-to-length (D/L) ratios of 2.3 or 0.6. Further comparisons using sky radiance data obtained by Mars Exploration Rover (MER) mission's Pancam instrument also derived similar results (Lemmon et al., 2004; Smith and Wolff, 2014).

All of these investigations have shown that light scattering by dust in Mars' atmosphere is consistent with non-spherical randomly oriented particles; however, the limited number of observations and the seasonal data coverage may only provide dust aerosol properties at the particular time and place of the observation. The objective of this work is to characterise Martian atmospheric dust scattering phase function using sky image data captured by the Mars Science Laboratory (MSL) engineering cameras and to contribute to previous studies by extending the results with observations for multiple seasons covering 4 Martian Years (MY 31 to 34).

The MSL mission have evaluated dust properties and its atmospheric loading at Gale Crater (4.6ºS; 137.4ºE) using different instrumentation (e.g., Lemmon et al. 2014; Smith et al., 2016; Vicente-Retortillo et al., 2017; McConnochie et al., 2017). Although not initially designed for scientific use, images retrieved by rover engineering cameras (navigation and hazard avoidance cameras) can be used as a complementary source of data for atmospheric studies (Soderblom et al. 2008; Smith and Wolff, 2014; Wolfe and Lemmon, 2015; Moores et al., 2015; Kloos et al., 2018). This study is a continuation of the work started in Chen-Chen et al. (2019), in which dust column optical depth and aerosol particle size were derived using MSL navigation cameras. In this case, the large field-of-view (FOV) offered by the hazard avoidance cameras, together with their capability to obtain simultaneous observations and their frequent use, make them suitable for studying dust light scattering properties at medium and large scattering angles.

This manuscript is structured as follows. In Section 2 the observation dataset is described and the processing details for the calibration and geometric reduction of MSL engineering camera images are provided. Section 3 describes the methodology used in this study, including the different aerosol models considered and the radiative transfer based retrieval procedure. In Section 4 the outcomes of this work are presented and discussed, together with the uncertainties and limitations of the method. Finally, in Section 5 a summary of the findings of this research and future prospects are given.

## 2. MSL engineering cameras observations

The MSL rover is equipped with a set of 12 engineering cameras: 4 navigation cameras (Navcam) and 8 hazard avoidance cameras (Hazcam). These cameras are build-to-print copies of MER engineering cameras and their objective is to provide guidance to the rover and support its operation during the drive across the surface. Navcams are located at the remote sensing mast and have a 45-degree FOV. The 8 Hazcams are fixed to the rover's chassis; they are located at the front (4) and rear (4) of the vehicle and have 124-degree square FOV optics (*fish-eye lens*). All imagers are equipped with a 1024x1024 pixel CCD detector and a broadband visible filter with an effective wavelength of 650 nm. For a complete description and technical specifications of these cameras we refer to Maki et al. (2012). The information related to the performance of the electronics and optics can be found in Maki et al. (2003).

2.1 Image sequences

Both MSL Hazcam and Navcam observation data were used in this study. In the case of Hazcam, we have taken advantage of their simultaneous front-rear pointing wide FOV imaging capability to retrieve the angular distribution of Martian sky brightness (**Figure 1, top**). For observations taken between local true solar time (LTST) 16:00 and 17:30, the corresponding solar elevation angle is approximately between 25º to 5º and the solar almucantar plane (circle of sky points with same elevation angle as the Sun) is contained within Hazcams' FOV. Depending on the rover's orientation and the surrounding topography, it is possible to retrieve the sky radiance as a function of the scattering angle with a 110º coverage and reaching up to 160º of scattering angle (maximum scattering angle in the solar almucantar plane is given by $\theta_{max} = 180^o - 2\varepsilon_{Sun}$, where $\varepsilon_{Sun}$ is the solar elevation angle). Therefore, the sampling of sky radiances along the solar almucantar direction was chosen for Hazcam observations. Part of the images had to be manually evaluated in order to discard unwanted contributions to the observed sky brightness curve (e.g., rover's chassis, robotic arm, rocks, and other scenery elements).

We have also considered for this study Navcam full sky-survey sequences (**Figure 1, bottom**). These datasets consist of multiple observations (usually 17 or 18 images) obtained in the early morning or afternoon in which the complete upper hemisphere was captured. The sky radiance as a function of the scattering angle was retrieved by sampling along the solar almucantar, in an analogous way as for Hazcam images. As the Navcam imagers are located at the rover's remote sensing mast, they have a more flexible pointing capability and the sky radiance curves were easily retrieved without the need of any further manual image analysis step. Only the possible intersections of the solar almucantar plane with *Aeolis Mons* (Mount Sharp) at the backscattering region, in the case of low Sun elevation observations, had to be taken into account when performing the data retrieval.

We show on **Figure 2**, for all the observation data retrieved along the solar almucantar, the contour plot of the sky radiance as a function of the scattering angle ($\theta$) and the solar longitude ($L_S$). It can be appreciated that the sky brightness intensity and its angular distribution function follows a seasonal variation similar to the one derived for the dust column optical depth (e.g., Lemmon et al., 2014, Smith et al., 2016). The first part of the year (aphelion season) is characterised for its low dust activity and atmospheric optical depth; which can be also identified in the sky radiance curves, that show a steeper drop in the radiance values during this period ($L_S$ ~ 70º to 140º) in the lateral scattering region ($\theta$ = 90º to 120º), when compared to a flatter curve present during the high dust loading season, centred on $L_S$ = 200º.

The complete list of observations is provided on **Table A1** (on the Appendix).

2.2 Photometric calibration and geometric reduction

The raw EDR image files used in this work were converted from their original 12-bit pixel DN into physical units of absolute radiance (W $m^{-2}$ $nm^{-1}$ $sr^{-1}$). Once radiometrically calibrated, a geometric reduction was performed using the CAHVOR(E) photogrammetric camera model system (Yakimovsky and Cunningham, 1978; Gennery, 2006) in order to assign to each pixel their corresponding values of elevation and azimuth with respect to the local site reference frame (Peters, 2016). A detailed description of the radiometric calibration and geometric reduction of MSL Navcam observations can be found in Chen-Chen et al. (2019). An identical procedure was applied to MSL Hazcam images for this study and the specific calibration parameters for these cameras are provided in **Table 1**. The procedure derived for the calibration of MSL engineering cameras is based on the methodology developed by Soderblom et al. (2008) for the same instruments on-board the MER mission.

In order to validate Hazcam's calibration parameters, multiple comparisons of Navcam and Hazcam calibrated images were performed. Observation-pairs with similar pointing and near in LTST were selected for different sols and the absolute radiance values for same scenery features (e.g., sky, ground, Mount Sharp) were compared. The results of this procedure showed average differences of less than 5% between both imagers. As previous MSL Navcam absolute radiance uncertainty was estimated of about 12% (Chen-Chen et al., 2019), for this study we have considered Hazcam absolute radiance uncertainty of about 17%.

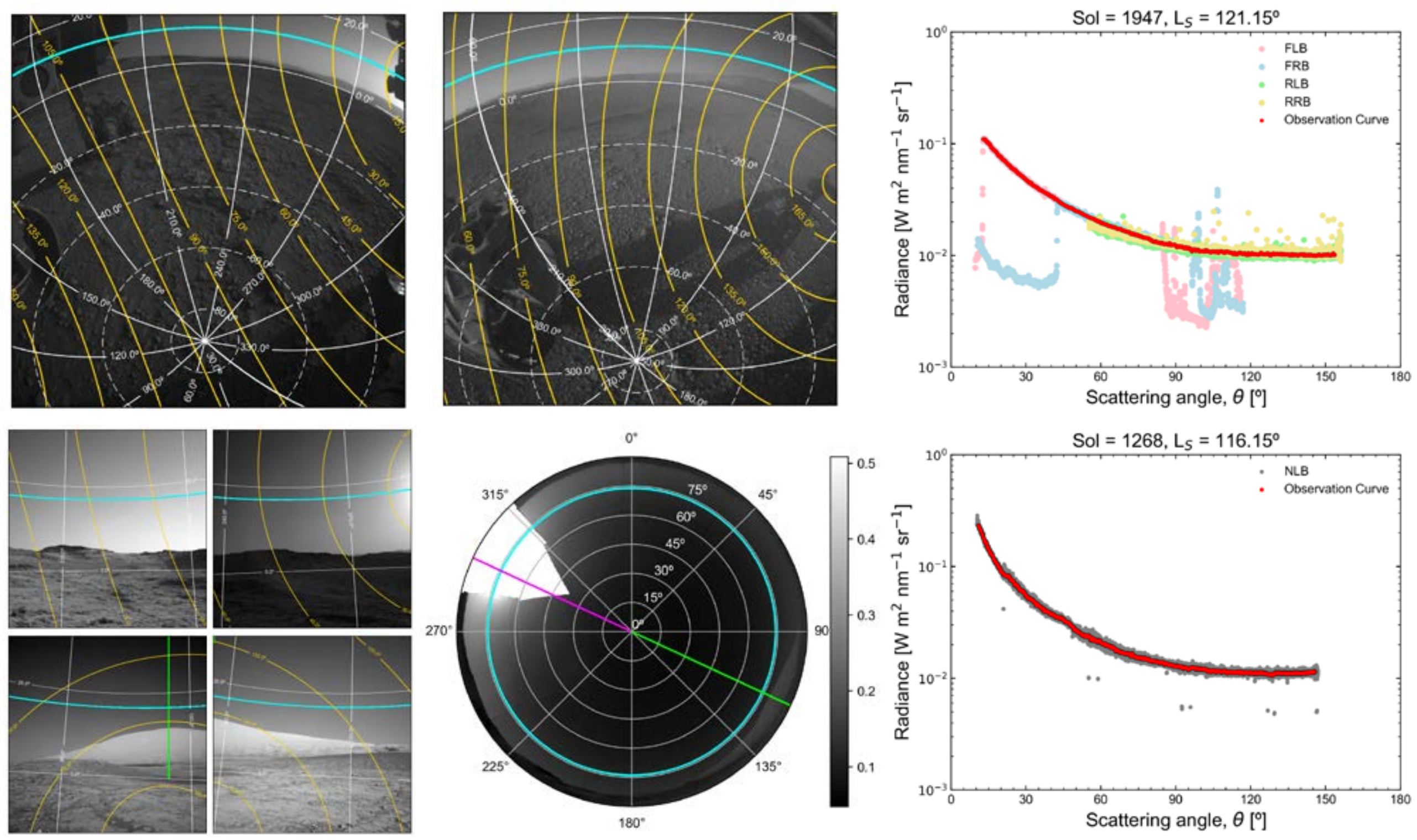

***Figure 1**. MSL engineering cameras observations. MSL Hazcam (top) and Navcam (bottom) images used for deriving the sky brightness angular distribution. The azimuth-elevation grid (white) with respect to a local site frame is shown, together with the scattering angles (yellow) and the solar almucantar plane (cyan). **Top:** Hazcam FLB (left) and RLB camera observations (centre) obtained on Sol 1947, $L_S$=121.15º, LTST ~ 17h, with solar elevation angle of 11º. (Right) The sky radiance retrieved by all Hazcam cameras (FLB, FRB, RLB, RRB) along the solar almucantar plane and the final observation curve derived from these contributions. **Bottom:** (Left) Observations part of the Navcam sky-survey sequence retrieved on Sol 1268, $L_S$=116.1º, LTST within 16:30 to 16:40, with Sun's elevation of 16º to 18º. (Centre) Polar-plot composition of the full sky-survey sequence, for clarity, the square root of radiance values is plotted. The almucantar (cyan) and solar principal plane's forward (magenta) and backward (green) region are also shown. On the right, the sky radiance sampled by each image of the Navcam sky-survey sequence on the solar almucantar (gray) and the final observation curve (red) are plotted. Additional data on the images are provided in Table A1 in the Appendix.*

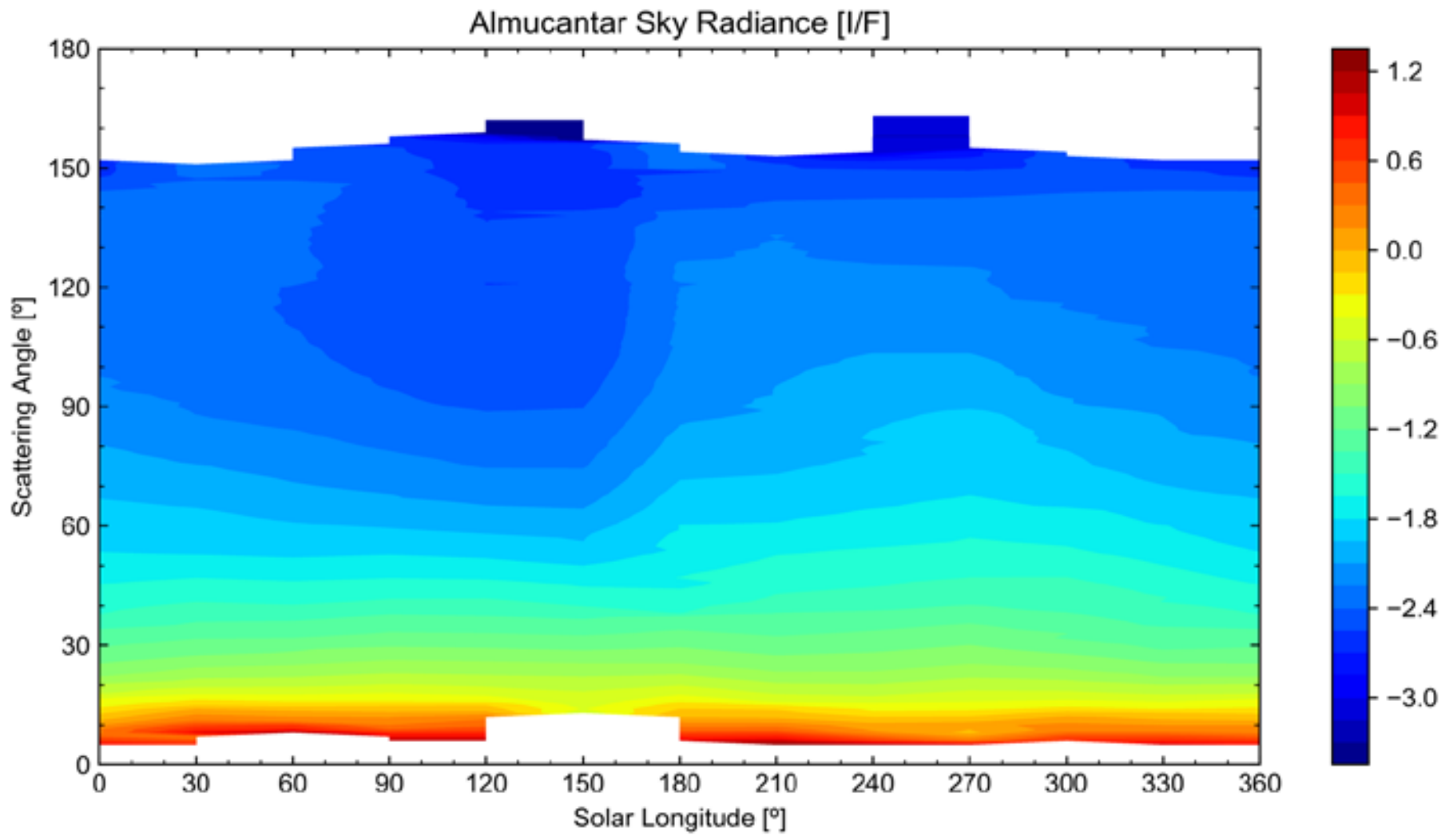

***Figure 2.** Seasonal variation of sky radiance as a function of the scattering angle. MSL engineering cameras retrieved sky brightness (expressed as radiance factor I/F, in logarithmic scale) in the solar almucantar plane, as a function of the scattering angle and the solar longitude. Radiance data binned every 1º of scattering angle and averaged over a 20º interval in $L_S$.*

Table 1. MSL Hazard Avoidance Cameras (Hazcam) calibration parameters

| CALIBRATION STAGE | MSL HAZARD AVOIDANCE CAMERAS CALIBRATION PARAMETERS (0) | | | | SOURCE |
|---|---|---|---|---|---|
| | **FHAZ_LEFT_B**: SN_0208 | **FHAZ_RIGHT_B:** SN_0209 | **RHAZ_LEFT_B:** SN_0212 | **RHAZ_RIGHT_B**: SN_0207 | |
| *Bias removal* | a0 = -25.45 DN, a1 = 36.5 DN, a2 = 0.0137 ºC$^{-1}$ | a0 = -42.92 DN, a1 = 56.7 DN, a2 = 0.0104 ºC$^{-1}$, | a0 = -10.78 DN, a1 = 26.1 DN, a2 = 0.0302 ºC$^{-1}$, | a0 = -2.12 DN, a1 = 29.3 DN, a2 = 0.0277 ºC$^{-1}$, | Derived for MSL |
| *Dark current removal: parameters* | Masked region mean rate: c0 = 9.976 DN; c1 = 0.0992 ºC$^{-1}$ Active region mean rate: d0 = 17.877 DN, d1 = 0.0928 ºC$^{-1}$ | | | | Derived for MSL, PDS (1) |
| *Masked dark flat image* | FLB_388221706EDR_F, FLB_388221713EDR_F | FRB_388221706EDR_F, FRB_388221713EDR_F | RLB_388221950EDR_F, RLB_388221958EDR_F | RRB_388221950EDR_F, RRB_388221958EDR_F | PDS (1)(2) |
| *Active dark flat image* | FLB_388221720EDR_F, FLB_388221830EDR_F | FRB_388221720EDR_F, FRB_388221830EDR_F | RLB_388221965EDR_F, RLB_388221993EDR_F | RRB_388221965EDR_F, RRB_388221993EDR_F | PDS (1)(2) |
| *Flat field correction* | MSL_FLAT_SN_0208.IMG | MSL_FLAT_SN_0209.IMG | MSL_FLAT_SN_0212.IMG | MSL_FLAT_SN_0207.IMG | PDS (3) |
| *Conversion to physical units* | K0 = 9.634e-6 W m$^2$ nm$^{-1}$ sr$^{-1}$ (DN s$^{-1}$)$^{-1}$ ; K1 = 1.035e-8 W m$^2$ nm$^{-1}$ sr$^{-1}$ (DN s$^{-1}$)$^{-1}$ ºC$^{-1}$ | | | | Adapted from MER (4) |

(0): Calibration parameters are defined in Chen-Chen et al. (2019) and references therein
(1): https://pds-imaging.jpl.nasa.gov/data/msl/MSLHAZ_0XXX/DATA/CRUISE/
(2): Dark masked and active flats are available in this public repository: http://www.ajax.ehu.es/hcc /
(3): https://pds-imaging.jpl.nasa.gov/data/msl/ MSLHAZ _0XXX/CALIB/
(4): Table 2 from Soderblom et al., (2008). A 15% of uncertainty is assumed for K0 and K1 values.

## 3. Model and methodology

In this work the radiative transfer modelled sky brightness curves were iteratively compared to MSL Engineering Camera observations, in order to derive the parameters of the dust single scattering phase function generating the best fitting simulation. In the next paragraphs we describe the methodology followed to model the angular distribution of sky radiance and the comparison criterion used.

### 3.1 Radiative transfer model

The radiative transfer equation was solved using the discrete ordinates method (Stamnes et al., 1988) for a multiple scattering plane-parallel atmosphere. Simulations were performed using a Python version (PyDISORT, Ádámkovics et al., 2016) of DISORT 2.1 with pseudo-spherical correction (CDISORT, Buras et al., 2011). The atmosphere was modelled with 30 layers distributed in linearly spaced pressure levels with a total height of 100 km. Atmospheric parameter inputs (pressure, temperature, density, etc.) were retrieved from the Mars Climate Database (Forget et al., 1999; Millour et al., 2015). The main constituent species of Mars' atmosphere considered in our model ($CO_2$, $H_2O$, $O_2$, $N_2$ and $O_3$) present no strong gas absorption within the MSL cameras wavelength range (600 to 800 nm), so their contribution to the opacity was considered negligible. For the Rayleigh scattering due to the $CO_2$, the scattering cross section was obtained from the model and constants in Sneep and Ubachs (2005). These assumptions are the same as those taken in Chen-Chen et al. (2019).

### 3.2 Aerosol model

The radiative transfer computations required only 3 parameters at each layer of the discretised atmosphere model for the radiance: the aerosol single scattering albedo ($\omega_0$), the single scattering phase function $P(\theta)$, and the vertical distribution of the aerosol optical depth, $\tau(z)$.

The dust optical depth at each layer $\tau(z)$ was modelled following a Conrath profile (Forget et al., 1999; Heavens et al., 2011), and the total column optical depth input value required in these profiles were retrieved from MSL Mastcam direct Sun imaging extinction measurements (Lemmon et al., 2014) and MSL Navcam retrievals (Chen-Chen et al., 2019).

For $\omega_0$ and $P(\theta)$, we selected for this study the following 3 modelling approaches:

- *Analytical phase function.* A set of analytical single scattering phase functions were generated using a Double Henyey-Greenstein (DHG) three-parameter analytical expression (Kattawar et al., 1975; Gillespie, 1992) in the form of:

$$P_{DHG}(\theta) = \alpha \frac{1-g_1{}^2}{(1-2g_1 cos\theta + g_1{}^2)^{3/2}} + (1-\alpha) \frac{1-g_2{}^2}{(1-2g_2 cos\theta + g_2{}^2)^{3/2}} \qquad (1)$$

Parameters controlling the forward scattering ($g_1$), backward scattering ($g_2$) and the forward-backward ratio ($\alpha$) were varied in order to simulate different aerosol phase functions (Ignatov, 1997; Zhang and Li, 2016). The $g_1$ parameter was iterated from 0.50 to 1.00 with steps of 0.01; $g_2$ was varied between –g1 and +g1 (50 divisions) in order to prevent the backward scattering lobe from being greater than forward lobe and to avoid negative phase function values (Zhang and Li, 2016). Finally, the parameter controlling the ratio ($\alpha$) was iterated from 0.50 to 1.00 (fully forward scattering case) with 0.01 steps. Again, this was set in order to control the overall shape of the phase function and use representatives of actual airborne dust phase functions (e.g., Mishchenko et al., 1997; Dubovik et al., 2006). The single scattering albedo was fixed to $\omega_0$ = 0.975 based on results derived from surface-orbit combined observations by Wolff et al. (2009) and particularised for MSL engineering cameras effective wavelength ($\lambda_{eff}$ ~ 650 nm)

- *T-Matrix*. Previous studies have indicated the need to take into account the non-sphericity when modelling the optical properties of Martian dust (e.g., Pollack et al, 1977; Chylek and Grams, 1978). Although there are available multiple models for calculating the scattering properties of non-spherical particles, software codes for simulating particle shapes with complex and irregular geometry or large ensembles of particles are very computationally demanding (Yurkin and Hoekstra, 2007; Wriedt, 2009). We selected the T-matrix code (https://www.giss.nasa.gov/staff/mmishchenko/t_matrix.html) (Mishchenko et al., 1998) to calculate the radiative properties of randomly oriented cylindrical particles with different diameter-to-length (D/L) aspect ratios and sizes. The shape was fixed to cylindrical particles as the calculated phase functions for single aspect radio cylinders do simulate well the usual airborne dust phase function in the lateral scattering region ($\theta$ approximately from 90º to 120º) (Mishchenko et al., 1997). This consideration avoided the need of introducing additional parameters associated to a distribution of aspect ratios when spheroidal particles are used, therefore reducing the number of comparisons to be performed and consequently the computation time of the retrieval. The aspect ratio parameter D/L was varied from 0.5 to 2.5, with steps of 0.1. The values of the single scattering albedo and phase function were calculated assuming a power law particle size distribution for volume equivalent effective radius ($r_{eff}$) varying from 0.10 to 1.70 µm in 0.02 µm steps (e.g., Chen-Chen et al., 2019), with effective variance $\nu_{eff}$ = 0.3 (e.g. Mishchenko et al., 1997; Dubovik et al., 2006); the refractive complex index was derived from Wolff et al. (2009).

- *Laboratory measurements of Martian dust analogues.* Experimental measurements of single scattering phase functions for different Martian dust analogue samples were retrieved from the Amsterdam-Granada Light Scattering database (https://www.iaa.csic.es/scattering/) (Muñoz et al., 2012). The scattering phase functions at 647 nm for basalt, JSC0, JSC200, JSC-1A and palagonite samples were evaluated in this study. For a comprehensive description regarding the properties of the samples, experimental set up and retrieval of the scattering matrices we refer to the corresponding publications: basalt, JSC0 and JSC200 (Dabrowska et al., 2015); JSC-1A (Escobar-Cerezo et al., 2018) and palagonite (Laan et al., 2009). The single scattering albedo for each sample was approximated using the Lorenz-Mie theory (Mishchenko et al., 1995). For these computations, the particle size distribution parameters ($r_{eff}$ , $\nu_{eff}$) and complex refractive index ($m$) of each sample were derived from the database: basalt ($r_{eff}$=6.9 µm, $\nu_{eff}$=7.0, $m$=1.52+$i$0.001), JSC0 ($r_{eff}$=29.5 µm, $\nu_{eff}$=1.1, $m$=1.5+$i$0.001), JSC200 ($r_{eff}$=28.1 µm, $\nu_{eff}$=1.2, $m$=1.5+$i$0.001), JSC-1A ($r_{eff}$=15.85 µm, $\nu_{eff}$=2.28, $m$=1.65+$i$0.003), palagonite ($r_{eff}$=4.5 µm, $\nu_{eff}$=7.3, $m$=1.52+$i$0.0005). The resulting single scattering albedos were: $\omega_{0,basalt}$= 0.892, $\omega_{0,JSC0}$= 0.701, $\omega_{0,JSC200}$= 0.633, $\omega_{0,JSC1A}$= 0.708 and $\omega_{0,palagonite}$= 0.960.

We summarise on **Table 2** the dust aerosol models used in this work and their related parameters.

### 3.3 Retrieval procedure

An iterative retrieval scheme was implemented based on the comparison of radiative transfer simulations and MSL engineering camera observations of Martian sky brightness as a function of the scattering angle. A lowest mean quadratic deviation $\chi^2$ criterion was considered for determining the best fitting curve.

For each Hazcam/Navcam observation:

1. Radiometric calibration and geometric reduction were performed as described in Section 2: for each image-pixel the corresponding values of absolute radiance, azimuth/elevation angles with respect to a Mars' local site reference system and the resulting scattering angle were calculated. The absolute radiance was then converted into approximated radiance factor (*I/F*) units by dividing each pixel's radiance value by the solar spectral irradiance at the top of the atmosphere at the time of the observation convolved to the Hazcam/Navcam bandpass (same for both imagers, 1.524 W $m^2$ $nm^{-1}$ $sr^{-1}$ at 1 AU) and divided by $\pi$ (e.g., Soderblom et al., 2008). The solar spectral irradiance data was obtained from Colina et al. (1996).

2. Retrieval of the observed sky brightness as a function of the scattering angle was performed by sampling radiance values along the solar almucantar plane.

3. The simulated sky brightness curves were generated using the radiative transfer model for different combinations of aerosol modelling parameters (**Table 2**) and allocated in a look-up-table (LUT).

5. The observed ( $I/F_{obs}(\theta_i)$ ) sky radiance angular distribution function and the modelled ( $I/F_{mod}(\theta_i)$ ) curves contained in the LUT were compared using a standard $\chi^2$ least squares quadratic error criterion:

$$\chi^2 = \sum_{i=1}^{N} \left( \frac{I/F_{obs}(\theta_i) - I/F_{mod}(\theta_i)}{\sigma_i \cdot I/F_{obs}(\theta_i)} \right)^2 \qquad (2)$$

For the N sampled points along the curve a variance of $\sigma_i$ = 0.20 was used as a conservative value associated to the absolute calibration uncertainty for MSL engineering cameras considered in Section 2. The reduced $\chi^2$ values ($\chi^2_{red}$) were calculated by dividing the obtained $\chi^2$ by the number of degrees of freedom $\boldsymbol{\nu} = N - f$, where $N$ is the number of sampled points and $f$ the number of free parameters in the retrieval ($f$ is equal to 3 for DHG, 2 for T-matrix and 1 for laboratory measurements) (**Table 2**)

6. The set of input parameters for each aerosol model generating the simulated sky brightness angular distribution with the minimum $\chi^2$ value was considered the solution of the retrieval (**Figure 3**). The uncertainty level of the solution was estimated from the 68% confidence region (1σ error).

Table 2. Aerosol model parameters for radiative transfer simulations

| **Aerosol model** | **Single scat. albedo, $\omega_0$** | **Phase function, P(θ)** | | |
|---|---|---|---|---|
| | | **Parameters** | **Range** | **Reference** |
| Double Henyey-Greenstein | 0.975 | Forward scattering ($g_1$), backward scattering ($g_2$), and ratio ($\alpha$) | $g_1$: 0.50 to 1.00, step of 0.01.<br>$g_2$: - $g_1$ to +$g_1$, 50 divisions.<br>α: 0.50 to 1.00, step of 0.01. | Zhang and Li, 2016 |
| T-Matrix | Calculated | Cylindrical particles: diameter-to-length aspect ratio (*D/L*), size distribution effective radius ($r_{eff}$) | *D/L*: 0.5 to 2.5, step of 0.1<br>$r_{eff}$: 0.10 to 1.70 µm, step of 0.02 | Mishchenko et al., 1998. |
| Laboratory measurements | Calculated (Lorenz-Mie) | Martian dust analogue sample experimental phase functions at 647 nm. | *Samples:*<br>Basalt, JSC0, JSC200, JSC-1A, Palagonite | Muñoz et al., 2012; Dabrowska et al. 2015; Escobar-Cerezo et al., 2018; Laan et al., 2009 |

## 4. Results and discussion

The methodology described in the previous Section was followed to retrieve the aerosol model parameters generating the best fitting sky radiance simulations. In this Section 4, the outcomes of the parameterisation scheme are presented. A discussion is provided for studying the seasonal behaviour and the interrelationships of the resulting parameters and the uncertainties of the retrieval are evaluated.

A summary table with the complete results of this study is provided on **Table A2** in the Appendix to this manuscript.

*Double Henyey-Greenstein phase function parameters.* The seasonal and interannual behaviour of the DHG analytical phase function parameters ($g_1$, $g_2$, $\alpha$), and their interrelationships are shown on the left and right column on **Figure 5**, respectively. The average values retrieved for each parameter are: $g_1$ = 0.889±0.098, $g_2$ = 0.094±0.250 and $\alpha$ = 0.743±0.106. When recurring to the expressions provided on Zhang and Li (2016), these parameter values generate a single scattering phase function with an asymmetry factor of $g$ = 0.687±0.081, which is in good agreement with previous results by Wolff et al. (2009) at the 650 nm effective wavelength of MSL engineering cameras. It can be appreciated on **Figure 5** that results for Hazcam observations (red) show a greater dispersion and larger uncertainties than Navcam dataset outcomes (blue). This is mainly related to the pointing particularities of each set of cameras; while mast-mounted free pointing Navcam sky-surveys are capable of retrieving sky radiance curves covering scattering angles from approximately 10º to 150º, rover chassis fixed Hazcam observations are highly dependent on the geometry configuration at the specific LTST and location, thus retrieving image-sets with very different scattering angle coverage.

Regarding the seasonal variability of the DHG parameters, the results obtained during the low opacity aphelion season ($L_S$ ~ 40º to 130º) show noticeable differences when compared to the rest of the year (the sensitivity to possible contribution from the aphelion cloud belt water-ice clouds in the retrieved sky radiance data during this particular season will be discussed below). In particular, the forward scattering parameter ($g_1$) values tend to be lower within this time. As phase function values in the forward scattering region ($\theta$ ~ 5º to 30º) are related to the size of the particle (e.g., Kaufman, 1994; Tomasko et al., 1999), this may suggest the detection of smaller dust particles during this season. However, due to the differences in the scattering angle coverage by each observation, the lack of data in the forward scattering region may originate part of the dispersion in the results, therefore not providing strong evidences for identifying any particular seasonal behaviour. Seasonal differences can be also appreciated in the backward scattering parameter $g_2$. In this case, the retrieved negative values are mostly located within the same aphelion period ($L_S$ ~ 40º to 130º). DHG analytical phase functions with a $g_2 < 0$ are featured with a positive slope at the end of the backscattering region (minimum of phase function is at $\theta$ < 180º, existence of a peak). However, as in the previous case, the existing dispersion in the retrieved data does not allow to identify a clear seasonal behaviour for this parameter.

The interrelationships between the DHG parameters are shown at the right column of **Figure 5**. In this case, output charts tend to be more clear and results show a positive correlation for $g_1 - g_2$ parameters, and negative correlations for $g_1 - \alpha$ and $g_2 - \alpha$, being more evident in the latter case. The obtained negative correlations points out the role of the parameter $\alpha$ as weighting factor for controlling the overall shape of the DHG phase function; when large lobes in the function are obtained at the forward scattering area ($g_1$ close to 1) or at the backscattering (negative $g_2$), the parameter $\alpha$ tends to balance the counterpart region by shifting to 0.5 or 1.0, respectively.

Finally, regarding the interannual variability analysis, the different number of available observation data per MY and its seasonal distribution, sums up to the abovementioned dispersion of the retrieval results. Therefore it is not possible to conclude that any particular interannual behaviour was derived from the evaluated data.

*Dust shape.* The retrieval results for the diameter-to-length aspect ratio parameter for randomly oriented cylindrical particles calculated with T-matrix are shown on **Figure 6**. The frequency of aspect ratio counts returned average *D/L* values of 0.70 and 1.90 with an uncertainty of about 0.20, when differentiating *D/L* values larger and smaller than 1.0. These results present a good agreement with previous studies (e.g.: 0.60 or 2.30 by Wolff et al. (2001)). Regarding the seasonal evolution of this parameter, although average values tend to be slightly larger when only considering the low opacity aphelion season, it is not possible to conclude that the retrieved results show any seasonal variability.

*Laboratory measurements of Martian dust analogues.* The results of the observation-model comparison retrieval showed that only two models generated the best fitting model curve: basalt (78% of the cases) and palagonite (22%). This outcome is mainly related to the significant differences that exist in the particle size distribution of the available dust analogue samples ($r_{eff,palagonite}$= 4.5 µm, $r_{eff,basalt}$= 6.9 µm, , $r_{eff,JSC1A}$= 15.85 µm, $r_{eff,JSC200}$= 28.1 µm, $r_{eff,JSC0}$= 29.5 µm), where it can be appreciated that the effective radius parameter for the remaining analogues are about an order of magnitude larger than the usual values reported for Martian atmospheric dust aerosol ($r_{eff}$ order of ~ 1 µm) by previous studies (Korablev et al., 2005; Smith, 2008; McConnochie et al., 2017; Chen-Chen et al., 2019). Previous studies comparing Martian airborne dust with experimental analogue measurements resulted in best fits to samples of palagonite (Clancy et al., 1995; Merikallio et al., 2013) and basalt (Dabrowksa et al., 2015). No relevant seasonal or interannual variability in the best fitting basalt or palagonite dust samples were found.

The sensitivity of the retrieved DHG parameters to variations of the input values for the single scattering albedo, dust column optical depth and possible presence of water-ice clouds during the aphelion season was evaluated by performing several simulations for these scenarios (**Figure 4**).

*Sensitivity to aerosol optical depth.* The atmospheric column optical depth is a required input parameter for radiative transfer simulations. Regular measurements from MSL Mastcam afternoon direct Sun-imaging (Lemmon et al., 2014) and MSL Navcam near Sun-pointing observations (Chen-Chen, et al., 2019) were used. Dust column optical depth values were interpolated at the observation's sol (or $L_S$ if there were no data available within a range of 20 sols), which could introduce some uncertainty in our retrieval procedure. The sensitivity of the results to uncertainties in column optical depth measurements was evaluated by simulating two scenarios containing 15% more and less dust atmospheric loading with respect to the nominal case. When the column optical depth was decreased, the analytical DHG phase function parameters g1, g2 and $\alpha$ showed a difference of about 4%, 5% and 2.5% respectively with respect to the base scenario; whereas in the case of an increment of the dust extinction the resulting differences were of the order of 2%, 9% and 3%.

*Sensitivity to single scattering albedo*. The simulated sky brightness also depended on the input value of dust single scattering albedo ($\omega_0$). As it has been abovementioned, for the case of analytical DHG phase functions the single scattering albedo was fixed to 0.975, which is a representative value for Martian dust (Wolff et al., 2009) at the effective wavelength of the cameras. The sensitivity of our retrieval procedure to variations in this parameter was evaluated by comparing the obtained results when the input $\omega_0$ was set to of 0.940 (e.g., Tomasko et al., 1999). The resulting output parameters $g_1$, $g_2$ and $\alpha$ varied in the order of 4%, 25% and 2%, respectively, with respect to the nominal scenario

*Sensitivity to presence of water-ice clouds*. Part of the observations used in this study were obtained during the aphelion season (centred on $L_s$ ~ 70º) and the possible presence of water-ice clouds from the aphelion cloud belt, developing around $L_s$ = 40º - 60º and dissipating near $L_s$ ~ 150º (e.g., Clancy et al., 1996, 2003; Madeleine et al., 2012) might introduce deviations in the dust phase function parameters retrieval. Although the majority of the observations were taken before 7h or after 16h (LTST), when detections of water-ice clouds are very low and the reported optical depth is almost negligible (Kloos et al., 2018), the sensitivity of the results to this phenomenon was evaluated. For an observation retrieved on sol 1132 ($L_s$ = 54.2º) corresponding to MY 33 (high cloud detection at Gale Crater, e.g. McConnochie et al., 2017; Kloos et al., 2018), a simulation was performed in which a water-ice cloud was added to the base model: the optical depth of the cloud was set to $\tau_{cloud}$ = 0.15 as a representative value of afternoon retrievals (Kloos et al., 2018), water-ice scattering properties $Q_{ext}$ and $\omega_0$ were derived from Warren (1984) and the single scattering phase function was modelled with an analytical DHG using water-ice representative parameters from Zhang and Li (2016). Differences between the simulated sky radiance as a function of the scattering angle for the base scenario and the water-ice cloud scenario were about 12% (lower than assumed uncertainty of 15%). When comparing with the observation for retrieving the parameters generating the best fitting curve, variations of the output $g_1$, $g_2$ and $\alpha$ parameters of the DHG analytical phase function were of about 4.0%, 4.5% and 15.0%, respectively. The resulting simulated sky radiance curve including a water-ice cloud model and dust phase function are provided in **Figure 4**.

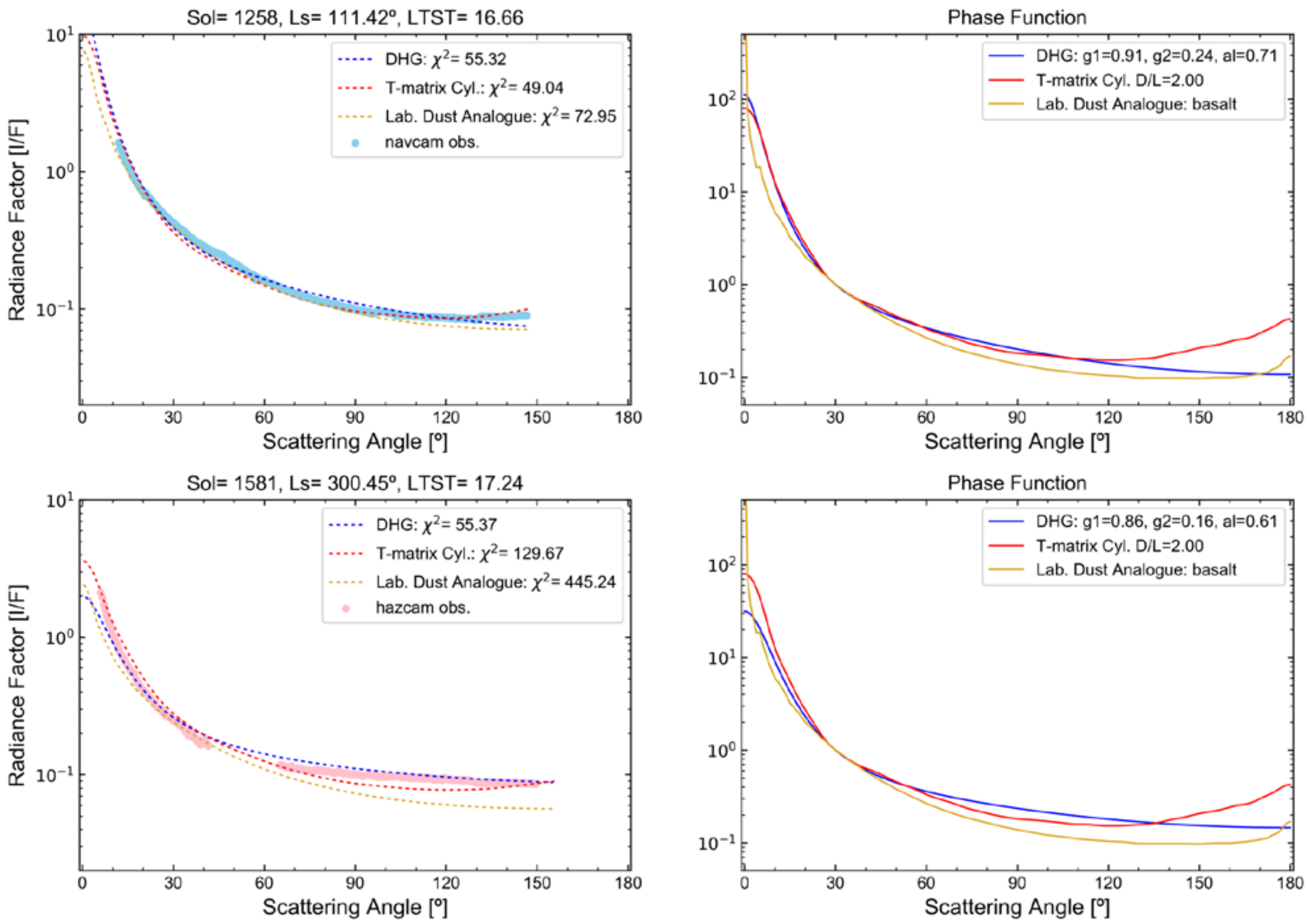

***Figure 3.*** *Comparison of observed and modelled sky brightness curves. Results of MSL Navcam (top) and Hazcam (bottom) observation comparisons to radiative transfer models: (left) the best fitting sky brightness as a function of the scattering angle simulations for the different aerosol models are provided; (right) the aerosol single scattering phase functions generating those best fitting curves. Phase functions are normalised to 1 at 30º scattering angle.*

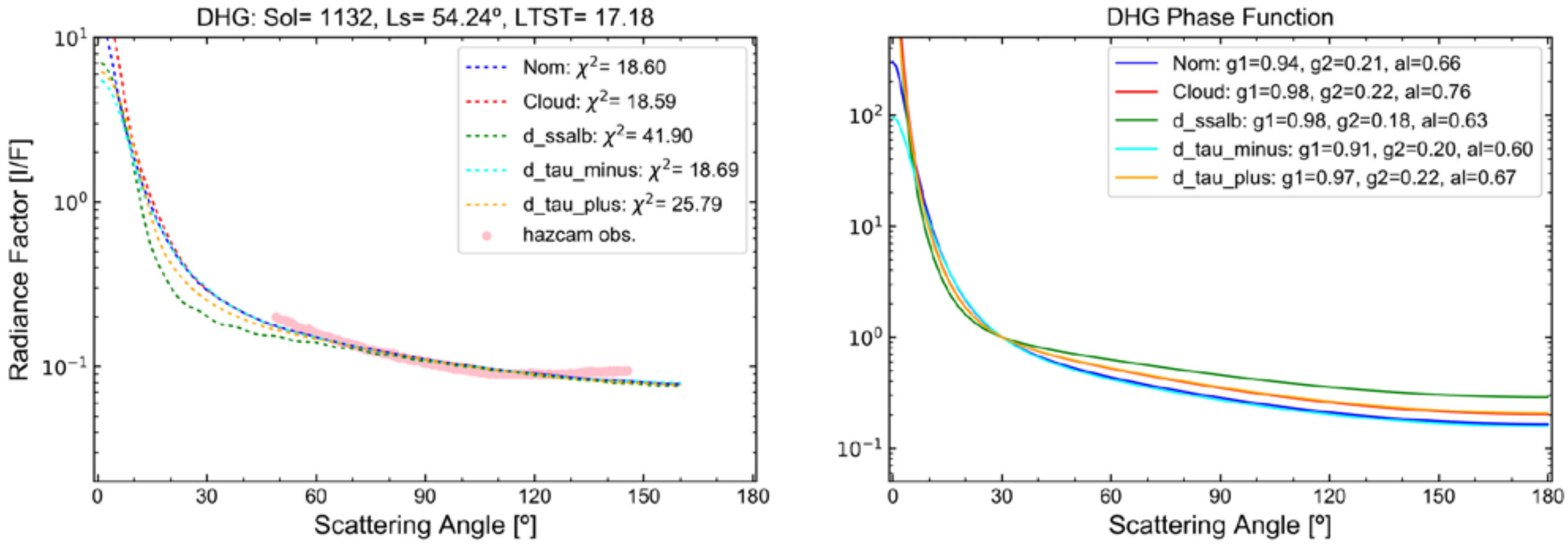

***Figure 4.*** *Sensitivity analysis of simulated sky brightness curves modelled with DHG aerosol phase functions. Outputs of DHG analytical phase function parameters ($g_1$, $g_2$, α) generating the best fitting sky brightness curve to MSL Hazcam observation corresponding to Sol 1132 (Ls = 54.24º, MY 33), under different simulation cases: nominal scenario (blue), presence of water-ice cloud (red), single scattering albedo set to ω0 = 0.94 (green), nominal dust column optical depth input value decreased 25% (cyan) and increased 25% (yellow). On the right, modelled sky radiance angular distribution compared to observation; left, DHG single scattering phase function curves generation those simulations. Phase functions are normalised to 1 at 30º scattering angle.*

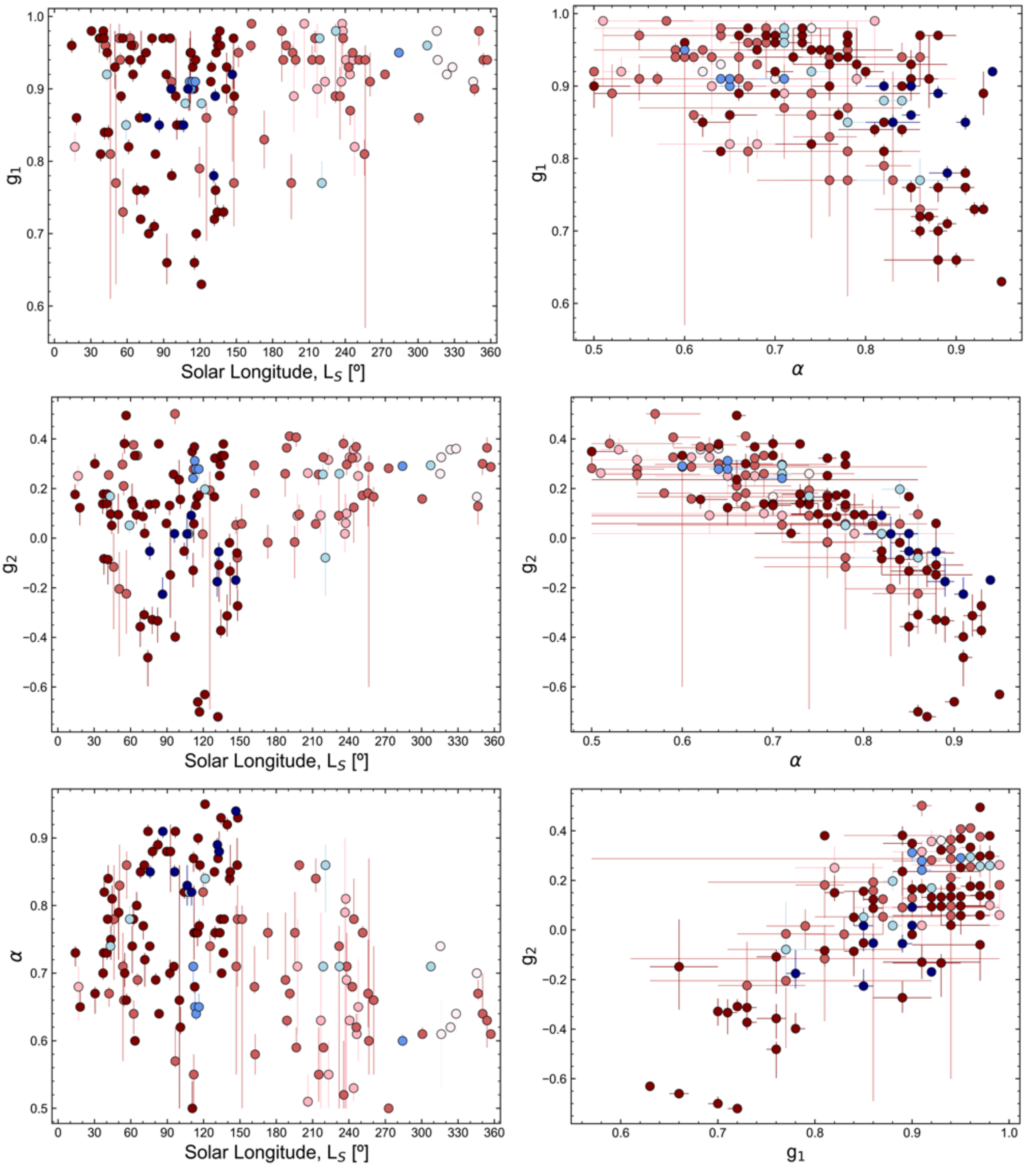

***Figure 5.*** *Double Henyey-Greenstein parameters seasonal variation and relationships. The seasonal behaviour (left column) and the existing interrelationships (right column) of the DHG analytical phase function parameters ($g_1$, $g_2$, α) generating the best fitting sky radiance model to MSL Navcam (blue) and Hazcam (red) observations. Colour shades indicate MY 31 (clearest) to MY 34 (darkest). No data for Navcam MY 31.*

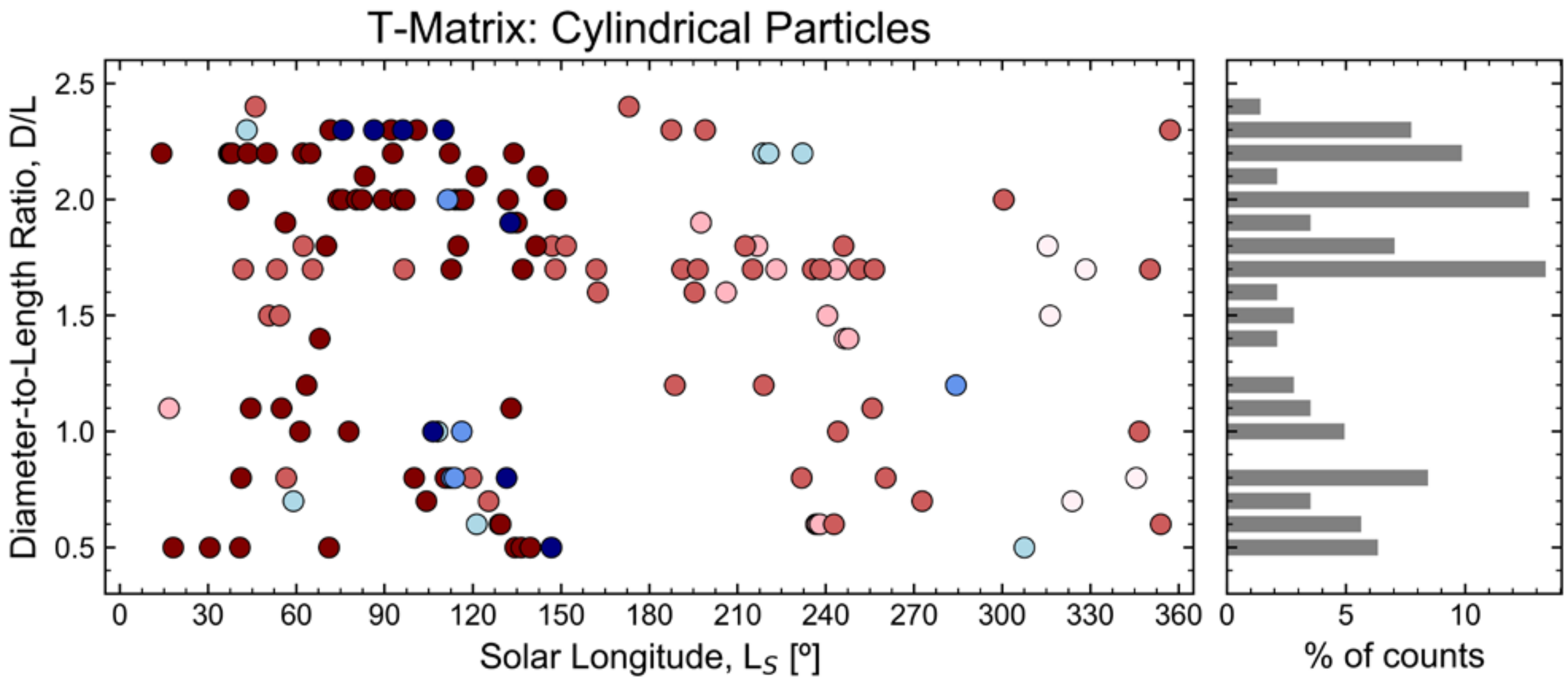

***Figure 6.*** *Seasonal and interannual variation of cylindrical particles aspect ratio. (Left) Results of T-matrix cylindrical particles diameter-to-length (D/L) aspect ratio parameter generating the best fitting sky radiance curve model to MSL Navcam (blue) and Hazcam (red) observations, as a function of the solar longitude and Martian Year (MY). Colour shades indicate MY 31 (clearest) to MY 34 (darkest). No data for Navcam MY 31. On the right, the bar chart shows the percentage of counts (frequency) for each D/L value.*

## 5. Conclusion

In this study we have used sky radiance measurements in the almucantar plane obtained by the MSL Engineering Cameras to constrain the Martian dust single scattering phase function. Hazcam simultaneous forward-rear pointing opportunistic afternoon observations and Navcam sky-survey image sequences were selected and photometric calibration and geometric reduction were performed on the raw images. The angular distribution of sky radiance was retrieved for different seasons and Martian Years. These observations contained data for the intermediate and large scattering angle region, from 30º up to about 160º, where the light scattering due to the aerosol is dominated by the shape of the particle.

The observed sky brightness curves were iteratively compared with radiative transfer sky radiance simulations. The modelled sky radiance as a function of the scattering angle were calculated following a parameterisation scheme for defining the dust single scattering phase function using 3 different aerosol modelling approaches: a three term Double Henyey-Greenstein analytical function, T-matrix code calculations for cylindrical particles and using experimental laboratory retrievals of Martian dust analogues.

Results retrieved from the comparison procedure show average Double Henyey-Greenstein parameter values of $g_1$ = 0.889±0.098, $g_2$ = 0.094±0.250, $\alpha$ = 0.743±0.106, which are related to a phase function with an asymmetry parameter of g = 0.673±0.081 (similar to, e.g., Wolff et al., 2009). Existing seasonal differences for the low dust opacity aphelion season ($L_S$ 30º to 150º) were observed for $g_1$ and $g_2$, although it was not possible to derive a clear seasonal or interannual behaviour, due mainly to the dispersion in the results and the different seasonal distribution of the data. Best fitting diameter-to-length aspect ratios for T-matrix cylindrical particles were of 0.70±0.20 and 1.90±0.20, presenting a good agreement with previous studies (Wolff et al., 2001). Comparisons with experimental single scattering phase functions of dust analogues returned only two different different best fitting samples, basalt (78%) and palagonite (22%), in line with Dabrowska et al., 2015.

Future research prospects include the retrieval and processing of observations under heavy dust loading scenarios, such as the global dust storm event in 2018, in order to evaluate the influence of such conditions in the dust single scattering phase function when compared to regular MY. On the aerosol modelling side, further developments can be done in the computation codes and methods for simulating the aerosol radiative properties for more realistic dust particles with complex shapes (e.g.; Yurkin and Hoekstra, 2007; Meng et al., 2007; Pitman et al., 2000). In addition to this, further comparisons can be performed for a broader variety of laboratory retrievals of Martian dust analogue single scattering measurements, with adequate particle size distributions, closer to the values retrieved for the atmospheric dust.

**Acknowledgements**

This work was supported by the Spanish project AYA2015-65041-P with FEDEER support, Grupos Gobierno Vasco IT-765-13, and Diputación Foral de Bizkaia - Aula EspaZio Gela. We wish to thank Professor Mark T. Lemmon for providing the MSL Mastcam optical depth values.

## APPENDIX A

Table A1. MSL Engineering Camera observations

| Sol | Ls[°] | MY | LTST | Sun Azim. [°] | Sun Elev. [°] | Camera | SCLK/Sequence |
|---|---|---|---|---|---|---|---|
| 269 | 315.37 | 31 | 7:12 | 106.81 | 18.60 | FLB, FRB,RLB, RRB | 421356422, 421356451 |
| 270 | 316.19 | 31 | 16:40 | 253.43 | 20.31 | FLB, FRB | 421480269 |
| 283 | 323.77 | 31 | 16:56 | 256.16 | 16.47 | FLB, FRB | 422635427 |
| 291 | 328.36 | 31 | 17:08 | 257.82 | 13.45 | FLB, FRB | 423346401 |
| 322 | 345.50 | 31 | 17:25 | 264.52 | 8.92 | FRB, FLB, RLB, RRB | 426099261, 426099296 |
| 383 | 16.67 | 32 | 16:48 | 278.81 | 17.12 | FRB, FLB, RLB, RRB | 431511077, 431511104 |
| 439 | 43.10 | 32 | 16:44 | 289.18 | 16.55 | NLB | ncam00548 |
| 474 | 59.00 | 32 | 16:52 | 293.41 | 13.90 | NLB | ncam00548 |
| 582 | 107.94 | 32 | 16:41 | 296.45 | 15.83 | NLB | ncam00548 |
| 610 | 121.23 | 32 | 17:34 | 291.85 | 4.25 | NLB | ncam00550 |
| 751 | 197.46 | 32 | 16:28 | 264.00 | 23.24 | FRB, FLB | 464174522 |
| 765 | 206.07 | 32 | 16:49 | 260.16 | 18.24 | FLB, FRB | 465418782 |
| 782 | 216.73 | 32 | 16:59 | 255.97 | 15.72 | FLB, FRB, RLB, RRB | 466928832, 466928866 |
| 785 | 218.61 | 32 | 16:10 | 255.03 | 27.69 | NLB | ncam00548 |
| 788 | 220.55 | 32 | 17:10 | 254.56 | 13.10 | NLB | ncam00548 |
| 792 | 223.09 | 32 | 16:31 | 253.61 | 22.58 | FLB, FRB, RLB, RRB | 467815011, 467815060 |
| 806 | 232.09 | 32 | 16:27 | 250.61 | 23.32 | NLB | ncam00548 |
| 813 | 236.64 | 32 | 17:09 | 249.67 | 13.49 | FLB, FRB | 469682237 |
| 814 | 237.29 | 32 | 16:55 | 249.46 | 16.70 | FLB, FRB | 469770196 |
| 815 | 237.93 | 32 | 16:46 | 249.23 | 18.71 | FLB, FRB | 469858472 |
| 819 | 240.53 | 32 | 16:45 | 248.58 | 18.91 | FLB, FRB | 470213649 |
| 824 | 243.77 | 32 | 16:02 | 246.89 | 28.79 | FLB, FRB | 470655026 |
| 828 | 246.40 | 32 | 16:46 | 247.31 | 18.68 | FLB, FRB | 471012999 |
| 830 | 247.49 | 32 | 16:28 | 246.77 | 22.74 | FLB, FRB | 471189532 |
| 924 | 307.50 | 32 | 16:21 | 250.39 | 24.74 | NLB | ncam00548 |
| 1105 | 41.93 | 33 | 17:11 | 287.74 | 10.24 | FLB, FRB | 495606448 |
| 1114 | 46.05 | 33 | 16:41 | 290.29 | 17.15 | FLB, FRB | 496403367 |
| 1124 | 50.60 | 33 | 16:44 | 291.54 | 16.13 | FLB, FRB | 497291116 |
| 1130 | 53.33 | 33 | 17:18 | 290.95 | 8.08 | FLB, FRB | 497825752 |
| 1132 | 54.24 | 33 | 17:10 | 291.47 | 9.92 | FLB, FRB | 498002768 |
| 1137 | 56.49 | 33 | 16:45 | 293.10 | 15.56 | FLB, FRB | 498445016 |
| 1150 | 62.35 | 33 | 17:14 | 293.29 | 8.78 | FLB, FRB | 499600628 |
| 1157 | 65.49 | 33 | 17:13 | 293.95 | 8.88 | FLB, FRB | 500221886 |
| 1226 | 96.60 | 33 | 16:59 | 296.75 | 11.70 | FLB, FRB | 506345674 |
| 1258 | 111.42 | 33 | 16:39 | 296.01 | 16.44 | NLB | ncam00548 |
| 1259 | 111.90 | 33 | 17:02 | 294.87 | 11.36 | FLB, FRB | 509275141 |
| 1261 | 112.82 | 33 | 16:01 | 298.19 | 25.02 | NLB | ncam00548 |
| 1263 | 113.77 | 33 | 16:18 | 296.84 | 21.38 | NLB | ncam00548 |
| 1268 | 116.15 | 33 | 16:36 | 295.27 | 17.30 | NLB | ncam00548 |
| 1275 | 119.51 | 33 | 17:12 | 292.95 | 9.33 | FLB, FRB | 510696014 |
| 1287 | 125.30 | 33 | 16:30 | 293.42 | 19.22 | FLB, FRB | 511758612 |
| 1330 | 146.96 | 33 | 17:16 | 284.46 | 9.51 | FLB, FRB | 515578339 |
| 1332 | 147.99 | 33 | 16:40 | 285.36 | 18.14 | FLB, FRB | 515753675 |
| 1339 | 151.68 | 33 | 17:09 | 282.87 | 11.25 | FLB, FRB | 516376834 |
| 1358 | 161.91 | 33 | 17:35 | 278.09 | 5.34 | FLB, FRB | 518064970 |
| 1359 | 162.44 | 33 | 17:03 | 278.73 | 13.41 | FLB, FRB | 518151723 |
| 1378 | 173.02 | 33 | 16:46 | 274.64 | 18.00 | FLB, FRB | 519837215 |
| 1403 | 187.49 | 33 | 16:15 | 268.73 | 26.17 | RLB, RRB | 522054677 |
| 1405 | 188.69 | 33 | 16:38 | 267.79 | 20.48 | FLB, FRB | 522233564 |
| 1409 | 191.09 | 33 | 17:24 | 265.98 | 9.14 | FLB, FRB, RLB, RRB | 522591532, 522591583 |
| 1416 | 195.28 | 33 | 16:38 | 264.86 | 20.72 | FLB, FRB | 523210121 |
| 1418 | 196.50 | 33 | 17:16 | 263.83 | 11.31 | FLB, FRB | 523390041 |
| 1422 | 198.93 | 33 | 17:05 | 262.97 | 14.10 | FLB, FRB, RLB, RRB | 523744454, 523744503 |
| 1444 | 212.57 | 33 | 17:18 | 257.39 | 11.12 | FLB, FRB, RLB, RRB | 525698534, 525698583 |
| 1448 | 215.07 | 33 | 16:53 | 256.60 | 17.22 | FLB, FRB | 526052099 |
| 1454 | 218.88 | 33 | 17:04 | 255.18 | 14.60 | FLB, FRB, RLB, RRB | 526585575, 526585624 |
| 1474 | 231.71 | 33 | 17:21 | 250.98 | 10.57 | FLB, FRB | 528361546 |
| 1480 | 235.57 | 33 | 16:37 | 249.75 | 20.95 | FLB, FRB | 528892687 |
| 1484 | 238.19 | 33 | 17:30 | 249.22 | 8.64 | FLB, FRB, RLB, RRB | 529251205, 529251256 |
| 1491 | 242.74 | 33 | 17:17 | 248.23 | 11.55 | FLB, FRB, RLB, RRB | 529872102, 529872144 |
| 1493 | 244.04 | 33 | 17:11 | 247.95 | 13.03 | FLB, FRB, RLB, RRB | 530049331, 530049373 |
| 1496 | 246.01 | 33 | 17:35 | 247.48 | 7.38 | FLB, FRB, RLB, RRB | 530317281, 530317325 |
| 1504 | 251.23 | 33 | 17:28 | 246.64 | 8.97 | FLB, FRB, RLB, RRB | 531027377, 531027427 |
| 1511 | 255.78 | 33 | 16:42 | 245.76 | 19.61 | FLB, FRB, RLB, RRB | 531646212, 531646260 |
| 1512 | 256.42 | 33 | 16:28 | 245.42 | 22.63 | FLB, FRB | 531734165 |
| 1518 | 260.35 | 33 | 17:02 | 245.53 | 15.07 | FLB, FRB | 532269165 |
| 1537 | 272.68 | 33 | 17:17 | 245.24 | 11.64 | FLB, FRB, RLB, RRB | 533957634, 533957668 |
| 1555 | 284.18 | 33 | 16:47 | 245.83 | 18.45 | NLB | ncam00548 |

| Sol | Ls[°] | MY | LTST | Sun Azim. [°] | Sun Elev. [°] | Camera | SCLK/Sequence |
|---|---|---|---|---|---|---|---|
| 1581 | 300.45 | 33 | 17:14 | 248.95 | 12.33 | FLB, FRB, RLB, RRB | 537865092, 537865126 |
| 1661 | 346.68 | 33 | 16:44 | 265.56 | 19.20 | FLB, FRB | 544965612 |
| 1668 | 350.22 | 33 | 17:15 | 266.69 | 11.29 | FLB, FRB, RLB, RRB | 545588880, 545588924 |
| 1675 | 353.89 | 33 | 16:57 | 268.62 | 15.86 | FLB, FRB, RLB, RRB | 546209039, 546209081 |
| 1681 | 357.01 | 33 | 16:46 | 270.20 | 18.40 | FLB, FRB, RLB, RRB | 546740938, 546740972 |
| 1715 | 14.15 | 34 | 16:34 | 278.18 | 20.70 | FLB, FRB | 549757840 |
| 1723 | 18.07 | 34 | 17:10 | 278.74 | 11.63 | FLB, FRB, RLB, RRB | 550470076, 550470111 |
| 1749 | 30.49 | 34 | 16:48 | 284.46 | 16.36 | FLB, FRB, RLB, RRB | 552776265, 552776300 |
| 1763 | 37.03 | 34 | 16:16 | 288.45 | 23.67 | FLB, FRB | 554016798 |
| 1764 | 37.50 | 34 | 16:46 | 287.15 | 16.40 | FLB, FRB, RLB, RRB | 554107432, 554107474 |
| 1765 | 37.97 | 34 | 17:16 | 286.21 | 9.12 | FLB, FRB, RLB, RRB | 554198056, 554198100 |
| 1770 | 40.27 | 34 | 16:18 | 289.50 | 22.85 | FLB, FRB | 554638225 |
| 1771 | 40.74 | 34 | 16:48 | 288.20 | 15.68 | FLB, FRB, RLB, RRB | 554728843, 554728878 |
| 1772 | 41.21 | 34 | 17:19 | 287.25 | 8.44 | FLB, FRB, RLB, RRB | 554819466, 554819510 |
| 1777 | 43.49 | 34 | 16:20 | 290.49 | 22.06 | FLB, FRB | 555259647 |
| 1779 | 44.42 | 34 | 17:21 | 288.22 | 7.71 | FLB, FRB, RLB, RRB | 555440886, 555440930 |
| 1791 | 49.88 | 34 | 16:31 | 291.98 | 19.26 | FLB, FRB | 556502839 |
| 1802 | 54.86 | 34 | 16:06 | 294.89 | 24.52 | FLB, FRB | 557477610 |
| 1805 | 56.23 | 34 | 16:38 | 293.36 | 17.14 | FLB, FRB | 557745889 |
| 1816 | 61.19 | 34 | 17:16 | 292.96 | 8.31 | FLB, FRB, RLB, RRB | 558724543, 558724579 |
| 1818 | 62.08 | 34 | 16:49 | 294.23 | 14.39 | FLB, FRB | 558901332 |
| 1821 | 63.42 | 34 | 16:28 | 295.59 | 19.01 | FLB, FRB, RLB, RRB | 559165415, 559165450 |
| 1824 | 64.78 | 34 | 16:55 | 294.52 | 12.89 | FLB, FRB, RLB, RRB | 559433352, 559433387 |
| 1831 | 67.92 | 34 | 17:07 | 294.62 | 10.12 | FLB, FRB, RLB, RRB | 560055379, 560055414 |
| 1836 | 70.17 | 34 | 17:18 | 294.58 | 7.53 | FLB, FRB, RLB, RRB | 560499871, 560499907 |
| 1838 | 71.06 | 34 | 17:01 | 295.39 | 11.49 | FLB, FRB, RLB, RRB | 560676312, 560676347 |
| 1839 | 71.49 | 34 | 16:09 | 298.29 | 22.90 | FLB, FRB, RLB, RRB | 560761909, 560761943 |
| 1845 | 74.19 | 34 | 16:41 | 296.73 | 15.71 | FLB, FRB, RLB, RRB | 561296467, 561296488 |
| 1848 | 75.53 | 34 | 16:19 | 298.17 | 20.55 | FLB, FRB, RLB, RRB | 561561352, 561561451 |
| 1849 | 75.81 | 34 | 07:27 | 62.56 | 17.74 | NRB | ncam00581 |
| 1853 | 77.79 | 34 | 16:54 | 296.51 | 12.74 | FLB, FRB, RLB, RRB | 562007347, 562007382 |
| 1859 | 80.47 | 34 | 16:19 | 298.67 | 20.44 | FLB, FRB, RLB, RRB | 562537763, 562537797 |
| 1863 | 82.28 | 34 | 16:48 | 297.18 | 14.11 | FLB, FRB, RLB, RRB | 562894592, 562894627 |
| 1865 | 83.18 | 34 | 16:42 | 297.55 | 15.46 | FLB, FRB, RLB, RRB | 563071738, 563071766 |
| 1872 | 86.34 | 34 | 17:01 | 296.80 | 11.25 | NRB | ncam00581 |
| 1879 | 89.51 | 34 | 17:02 | 296.76 | 10.81 | FLB, FRB, RLB, RRB | 564315716, 564315760 |
| 1885 | 92.21 | 34 | 16:05 | 299.99 | 23.33 | FLB, FRB | 564844795 |
| 1886 | 92.68 | 34 | 16:59 | 296.89 | 11.61 | FLB, FRB, RLB, RRB | 564936854, 564936898 |
| 1892 | 95.41 | 34 | 16:37 | 297.87 | 16.48 | FLB, FRB, RLB, RRB | 565468106, 565468141 |
| 1894 | 96.14 | 34 | 07:12 | 62.66 | 14.36 | NRB | ncam00581 |
| 1895 | 96.78 | 34 | 16:37 | 297.78 | 16.40 | FLB, FRB, RLB, RRB | 565734434, 565734468 |
| 1902 | 100.00 | 34 | 17:12 | 295.98 | 8.73 | FLB, FRB | 566357944 |
| 1904 | 100.92 | 34 | 17:04 | 296.23 | 10.62 | FLB, FRB, RLB, RRB | 566534944, 566534987 |
| 1911 | 104.16 | 34 | 17:16 | 295.43 | 7.90 | FLB, FRB, RLB, RBB | 567157071, 567157100 |
| 1916 | 106.47 | 34 | 16:27 | 297.47 | 19.00 | NRB | ncam00582 |
| 1924 | 110.02 | 34 | 06:59 | 64.76 | 11.59 | NRB | ncam00581 |
| 1925 | 110.69 | 34 | 17:06 | 294.91 | 10.39 | FLB, FRB, RLB, RRB | 568399163, 568399191 |
| 1927 | 111.62 | 34 | 16:53 | 295.26 | 13.18 | FLB, FRB | 568575951 |
| 1928 | 112.08 | 34 | 16:20 | 297.01 | 20.77 | FLB, FRB, RLB, RRB | 568662583, 568662679 |
| 1929 | 112.58 | 34 | 17:31 | 293.73 | 4.75 | RLB, RRB | 568755797 |
| 1932 | 113.98 | 34 | 16:50 | 294.99 | 14.01 | FLB, FRB, RLB, RRB | 569019566, 569019601 |
| 1934 | 114.94 | 34 | 17:23 | 293.52 | 6.51 | RLB, RRB | 569199181 |
| 1935 | 115.41 | 34 | 16:53 | 294.59 | 13.45 | FLB, FRB, RLB, RRB | 569286026, 569286061 |
| 1937 | 116.36 | 34 | 16:53 | 294.40 | 13.51 | FLB, FRB, RLB, RRB | 569463546, 569463588 |
| 1938 | 116.83 | 34 | 16:57 | 294.11 | 12.56 | RLB, RRB | 569552596 |
| 1947 | 121.15 | 34 | 17:04 | 292.86 | 11.12 | FLB, FRB, RLB, RRB | 570351901, 570351944 |
| 1963 | 128.95 | 34 | 17:27 | 290.05 | 6.18 | FLB, FRB, RLB, RRB | 571773544, 571773587 |
| 1964 | 129.44 | 34 | 17:18 | 290.18 | 8.25 | FLB, FRB, RLB, RRB | 571861758, 571861809 |
| 1968 | 131.40 | 34 | 16:23 | 292.00 | 21.03 | NRB | ncam00581 |
| 1969 | 131.91 | 34 | 17:05 | 289.92 | 11.42 | FLB, FRB | 572304802 |
| 1971 | 132.70 | 34 | 07:12 | 69.62 | 15.48 | NRB | ncam00583 |
| 1971 | 132.91 | 34 | 17:14 | 289.28 | 9.28 | FLB, FRB, RLB, RRB | 572482889, 572482940 |
| 1972 | 133.40 | 34 | 16:45 | 290.26 | 16.04 | FLB, FRB, RLB, RRB | 572569883, 572569932 |
| 1973 | 133.91 | 34 | 17:03 | 289.36 | 11.86 | FLB, FRB, RLB, RRB | 572659745, 572659794 |
| 1974 | 134.41 | 34 | 17:12 | 288.90 | 9.90 | FLB, FRB, RLB, RRB | 572749030, 572749080 |
| 1975 | 134.90 | 34 | 16:39 | 290.06 | 17.55 | FLB, FRB, RLB, RRB | 572835795, 572835844 |
| 1978 | 136.41 | 34 | 16:49 | 289.11 | 15.28 | FLB, FRB | 573102731 |
| 1979 | 136.93 | 34 | 17:25 | 287.67 | 6.87 | FLB, FRB, RLB, RRB | 573193678, 573193723 |
| 1984 | 139.46 | 34 | 17:10 | 287.29 | 10.45 | FLB, FRB, RLB, RRB | 573636601, 573636651 |
| 1988 | 141.50 | 34 | 17:29 | 286.03 | 6.15 | FLB, FRB, RLB, RRB | 573992786, 573992829 |
| 1989 | 142.00 | 34 | 16:30 | 288.08 | 20.29 | FLB, FRB, RLB, RRB | 574077879, 574077965 |
| 1998 | 146.66 | 34 | 17:03 | 284.98 | 12.47 | NRB | ncam00581 |
| 2000 | 147.69 | 34 | 16:49 | 285.12 | 16.02 | FLB, FRB, RLB, RRB | 575055503, 575055544 |
| 2001 | 148.22 | 34 | 17:05 | 284.32 | 12.04 | FLB, FRB, RLB, RRB | 575145295, 575145330 |

<u>Table A2. Results of the retrieval</u>

| Sol | Ls[°] | MY | LTST | Sun Elev [°] | Cam | DHG, $g_1$ | DHG, $g_2$ | DHG, α | CYL, D/L | TAB, Dust Analog. | $X^2_{red}$ DHG | $X^2_{red}$ CYL | $X^2_{red}$ TAB |
|---|---|---|---|---|---|---|---|---|---|---|---|---|---|
| 269 | 315.37 | 31 | 7:12 | 18.60 | HAZ | $0.98^{+0.00}_{-0.00}$ | $0.26^{+0.00}_{-0.00}$ | $0.74^{+0.01}_{-0.01}$ | 1.8 | Palagonite | 0.022 | 0.116 | 2.249 |
| 270 | 316.19 | 31 | 16:40 | 20.31 | HAZ | $0.94^{+0.03}_{-0.02}$ | $0.33^{+0.05}_{-0.04}$ | $0.61^{+0.05}_{-0.08}$ | 1.5 | Palagonite | 0.048 | 0.010 | 2.199 |
| 283 | 323.77 | 31 | 16:56 | 16.47 | HAZ | $0.92^{+0.02}_{-0.01}$ | $0.36^{+0.01}_{-0.00}$ | $0.62^{+0.00}_{-0.01}$ | 0.7 | Basalt | 0.013 | 0.011 | 2.784 |
| 291 | 328.36 | 31 | 17:08 | 13.45 | HAZ | $0.93^{+0.01}_{-0.00}$ | $0.36^{+0.04}_{-0.00}$ | $0.64^{+0.00}_{-0.03}$ | 1.7 | Basalt | 0.017 | 0.044 | 4.000 |
| 322 | 345.50 | 31 | 17:25 | 8.92 | HAZ | $0.91^{+0.00}_{-0.00}$ | $0.17^{+0.00}_{-0.00}$ | $0.70^{+0.00}_{-0.00}$ | 0.8 | Basalt | 0.099 | 0.024 | 4.450 |
| 383 | 16.67 | 32 | 16:48 | 17.12 | HAZ | $0.82^{+0.02}_{-0.02}$ | $0.25^{+0.04}_{-0.04}$ | $0.68^{+0.03}_{-0.05}$ | 1.1 | Basalt | 0.043 | 0.144 | 1.212 |
| 439 | 43.10 | 32 | 16:44 | 16.55 | NAV | $0.92^{+0.00}_{-0.01}$ | $0.17^{+0.00}_{-0.04}$ | $0.74^{+0.02}_{-0.00}$ | 2.3 | Basalt | 0.049 | 0.093 | 1.163 |
| 474 | 59.00 | 32 | 16:52 | 13.90 | NAV | $0.85^{+0.01}_{-0.02}$ | $0.05^{+0.00}_{-0.04}$ | $0.78^{+0.02}_{-0.00}$ | 0.7 | Basalt | 0.042 | 0.025 | 0.948 |
| 582 | 107.94 | 32 | 16:41 | 15.83 | NAV | $0.88^{+0.01}_{-0.00}$ | $0.02^{+0.04}_{-0.00}$ | $0.82^{+0.00}_{-0.01}$ | 1.0 | Basalt | 0.042 | 0.077 | 0.420 |
| 610 | 121.23 | 32 | 17:34 | 4.25 | NAV | $0.88^{+0.00}_{-0.01}$ | $0.20^{+0.04}_{-0.04}$ | $0.84^{+0.01}_{-0.02}$ | 0.6 | Palagonite | 0.039 | 0.296 | 0.365 |
| 751 | 197.46 | 32 | 16:28 | 23.24 | HAZ | $0.89^{+0.09}_{-0.05}$ | $0.09^{+0.09}_{-0.11}$ | $0.71^{+0.07}_{-0.10}$ | 1.9 | Basalt | 0.002 | 0.026 | 1.491 |
| 765 | 206.07 | 32 | 16:49 | 18.24 | HAZ | $0.99^{+0.00}_{-0.09}$ | $0.26^{+0.00}_{-0.02}$ | $0.51^{+0.04}_{-0.01}$ | 1.6 | Basalt | 0.031 | 0.025 | 2.721 |
| 782 | 216.73 | 32 | 16:59 | 15.72 | HAZ | $0.90^{+0.09}_{-0.04}$ | $0.09^{+0.08}_{-0.07}$ | $0.63^{+0.16}_{-0.13}$ | 1.8 | Basalt | 0.002 | 0.010 | 4.529 |
| 785 | 218.61 | 32 | 16:10 | 27.69 | NAV | $0.97^{+0.00}_{-0.00}$ | $0.26^{+0.04}_{-0.00}$ | $0.71^{+0.01}_{-0.01}$ | 2.2 | Basalt | 0.036 | 0.286 | 0.601 |
| 788 | 220.55 | 32 | 17:10 | 13.10 | NAV | $0.77^{+0.03}_{-0.01}$ | $-0.08^{+0.19}_{-0.15}$ | $0.86^{+0.03}_{-0.07}$ | 2.2 | Basalt | 0.035 | 0.113 | 3.417 |
| 792 | 223.09 | 32 | 16:31 | 22.58 | HAZ | $0.91^{+0.02}_{-0.03}$ | $0.32^{+0.04}_{-0.04}$ | $0.55^{+0.05}_{-0.05}$ | 1.7 | Basalt | 0.051 | 0.011 | 2.867 |
| 806 | 232.09 | 32 | 16:27 | 23.32 | NAV | $0.98^{+0.00}_{-0.00}$ | $0.26^{+0.00}_{-0.00}$ | $0.71^{+0.01}_{-0.00}$ | 2.2 | Basalt | 0.069 | 0.364 | 1.208 |
| 813 | 236.64 | 32 | 17:09 | 13.49 | HAZ | $0.91^{+0.08}_{-0.01}$ | $0.02^{+0.16}_{-0.08}$ | $0.79^{+0.11}_{-0.29}$ | 0.6 | Palagonite | 0.008 | 0.008 | 2.025 |
| 814 | 237.29 | 32 | 16:55 | 16.70 | HAZ | $0.99^{+0.00}_{-0.07}$ | $0.06^{+0.12}_{-0.00}$ | $0.81^{+0.00}_{-0.31}$ | 0.6 | Palagonite | 0.005 | 0.014 | 3.793 |
| 815 | 237.93 | 32 | 16:46 | 18.71 | HAZ | $0.98^{+0.01}_{-0.05}$ | $0.10^{+0.04}_{-0.04}$ | $0.69^{+0.06}_{-0.19}$ | 0.6 | Palagonite | 0.001 | 0.005 | 4.958 |
| 819 | 240.53 | 32 | 16:45 | 18.91 | HAZ | $0.94^{+0.01}_{-0.01}$ | $0.33^{+0.00}_{-0.00}$ | $0.63^{+0.01}_{-0.00}$ | 1.5 | Palagonite | 0.090 | 0.015 | 3.112 |
| 824 | 243.77 | 32 | 16:02 | 28.79 | HAZ | $0.92^{+0.02}_{-0.00}$ | $0.36^{+0.05}_{-0.00}$ | $0.53^{+0.01}_{-0.02}$ | 1.7 | Basalt | 0.020 | 0.169 | 1.369 |
| 828 | 246.40 | 32 | 16:46 | 18.68 | HAZ | $0.94^{+0.01}_{-0.01}$ | $0.33^{+0.00}_{-0.04}$ | $0.61^{+0.03}_{-0.00}$ | 1.4 | Basalt | 0.073 | 0.018 | 3.225 |
| 830 | 247.49 | 32 | 16:28 | 22.74 | HAZ | $0.82^{+0.05}_{-0.04}$ | $0.25^{+0.09}_{-0.08}$ | $0.96^{+0.00}_{-0.00}$ | 1.4 | Basalt | 0.005 | 0.055 | 2.075 |
| 924 | 307.50 | 32 | 16:21 | 24.74 | NAV | $0.96^{+0.00}_{-0.00}$ | $0.29^{+0.00}_{-0.00}$ | $0.71^{+0.00}_{-0.00}$ | 0.5 | Basalt | 0.101 | 0.178 | 0.113 |
| 1105 | 41.93 | 33 | 17:11 | 10.24 | HAZ | $0.96^{+0.00}_{-0.00}$ | $0.25^{+0.00}_{-0.00}$ | $0.67^{+0.01}_{-0.02}$ | 1.7 | Palagonite | 0.123 | 0.008 | 2.278 |
| 1114 | 46.05 | 33 | 16:41 | 17.15 | HAZ | $0.81^{+0.18}_{-0.20}$ | $-0.12^{+0.14}_{-0.25}$ | $0.78^{+0.08}_{-0.09}$ | 2.4 | Basalt | 0.003 | 0.012 | 2.844 |
| 1124 | 50.60 | 33 | 16:44 | 16.13 | HAZ | $0.77^{+0.15}_{-0.14}$ | $-0.20^{+0.15}_{-0.27}$ | $0.83^{+0.06}_{-0.05}$ | 1.5 | Basalt | 0.003 | 0.021 | 2.395 |
| 1130 | 53.33 | 33 | 17:18 | 8.08 | HAZ | $0.97^{+0.02}_{-0.00}$ | $0.30^{+0.05}_{-0.00}$ | $0.71^{+0.05}_{-0.14}$ | 1.7 | Palagonite | 0.021 | 0.168 | 1.879 |
| 1132 | 54.24 | 33 | 17:10 | 9.92 | HAZ | $0.94^{+0.01}_{-0.00}$ | $0.21^{+0.00}_{-0.04}$ | $0.66^{+0.06}_{-0.01}$ | 1.5 | Palagonite | 0.097 | 0.005 | 2.773 |
| 1137 | 56.49 | 33 | 16:45 | 15.56 | HAZ | $0.73^{+0.06}_{-0.03}$ | $-0.22^{+0.18}_{-0.16}$ | $0.86^{+0.02}_{-0.05}$ | 0.8 | Basalt | 0.005 | 0.007 | 1.289 |
| 1150 | 62.35 | 33 | 17:14 | 8.78 | HAZ | $0.97^{+0.01}_{-0.00}$ | $0.38^{+0.00}_{-0.00}$ | $0.64^{+0.02}_{-0.03}$ | 1.8 | Palagonite | 0.068 | 0.073 | 1.743 |
| 1157 | 65.49 | 33 | 17:13 | 8.88 | HAZ | $0.96^{+0.00}_{-0.01}$ | $0.33^{+0.00}_{-0.00}$ | $0.69^{+0.01}_{-0.01}$ | 1.7 | Palagonite | 0.144 | 0.074 | 2.272 |
| 1226 | 96.60 | 33 | 16:59 | 11.70 | HAZ | $0.91^{+0.01}_{-0.01}$ | $0.50^{+0.01}_{-0.04}$ | $0.57^{+0.05}_{-0.01}$ | 1.7 | Basalt | 0.147 | 0.129 | 2.273 |
| 1258 | 111.42 | 33 | 16:39 | 16.44 | NAV | $0.91^{+0.01}_{-0.00}$ | $0.24^{+0.04}_{-0.00}$ | $0.71^{+0.00}_{-0.03}$ | 2.0 | Basalt | 0.207 | 0.183 | 0.271 |
| 1259 | 111.90 | 33 | 17:02 | 11.36 | HAZ | $0.91^{+0.02}_{-0.03}$ | $0.28^{+0.04}_{-0.01}$ | $0.55^{+0.03}_{-0.05}$ | 0.8 | Basalt | 0.117 | 0.014 | 2.067 |
| 1261 | 112.82 | 33 | 16:01 | 25.02 | NAV | $0.90^{+0.00}_{-0.00}$ | $0.31^{+0.00}_{-0.04}$ | $0.65^{+0.02}_{-0.01}$ | 0.8 | Basalt | 0.189 | 0.108 | 0.166 |
| 1263 | 113.77 | 33 | 16:18 | 21.38 | NAV | $0.91^{+0.00}_{-0.00}$ | $0.28^{+0.00}_{-0.00}$ | $0.64^{+0.00}_{-0.01}$ | 0.8 | Basalt | 0.194 | 0.102 | 0.239 |
| 1268 | 116.15 | 33 | 16:36 | 17.30 | NAV | $0.91^{+0.00}_{-0.01}$ | $0.28^{+0.00}_{-0.04}$ | $0.65^{+0.02}_{-0.01}$ | 1.0 | Basalt | 0.191 | 0.073 | 0.258 |
| 1275 | 119.51 | 33 | 17:12 | 9.33 | HAZ | $0.79^{+0.03}_{-0.04}$ | $0.02^{+0.07}_{-0.09}$ | $0.82^{+0.03}_{-0.02}$ | 0.8 | Basalt | 0.055 | 0.013 | 1.154 |
| 1287 | 125.30 | 33 | 16:30 | 19.22 | HAZ | $0.86^{+0.02}_{-0.17}$ | $0.19^{+0.08}_{-0.88}$ | $0.74^{+0.02}_{-0.04}$ | 0.7 | Basalt | 0.015 | 0.017 | 0.198 |
| 1330 | 146.96 | 33 | 17:16 | 9.51 | HAZ | $0.87^{+0.06}_{-0.07}$ | $0.05^{+0.03}_{-0.04}$ | $0.71^{+0.09}_{-0.16}$ | 1.8 | Basalt | 0.013 | 0.026 | 3.062 |
| 1332 | 147.99 | 33 | 16:40 | 18.14 | HAZ | $0.77^{+0.12}_{-0.06}$ | $-0.08^{+0.13}_{-0.11}$ | $0.78^{+0.05}_{-0.10}$ | 1.7 | Basalt | 0.008 | 0.016 | 2.161 |
| 1339 | 151.68 | 33 | 17:09 | 11.25 | HAZ | $0.95^{+0.01}_{-0.02}$ | $0.06^{+0.08}_{-0.00}$ | $0.78^{+0.02}_{-0.28}$ | 1.8 | Palagonite | 0.021 | 0.027 | 3.663 |
| 1358 | 161.91 | 33 | 17:35 | 5.34 | HAZ | $0.96^{+0.00}_{-0.01}$ | $0.29^{+0.00}_{-0.04}$ | $0.68^{+0.06}_{-0.01}$ | 1.7 | Palagonite | 0.014 | 0.592 | 1.154 |
| 1359 | 162.44 | 33 | 17:03 | 13.41 | HAZ | $0.99^{+0.00}_{-0.00}$ | $0.18^{+0.00}_{-0.00}$ | $0.58^{+0.03}_{-0.03}$ | 1.6 | Basalt | 0.038 | 0.026 | 3.886 |
| 1378 | 173.02 | 33 | 16:46 | 18.00 | HAZ | $0.83^{+0.04}_{-0.04}$ | $-0.02^{+0.07}_{-0.07}$ | $0.76^{+0.03}_{-0.04}$ | 2.4 | Basalt | 0.010 | 0.031 | 1.861 |
| 1403 | 187.49 | 33 | 16:15 | 26.17 | HAZ | $0.98^{+0.01}_{-0.07}$ | $0.26^{+0.00}_{-0.09}$ | $0.69^{+0.10}_{-0.00}$ | 2.3 | Basalt | 0.010 | 0.042 | 0.038 |
| 1405 | 188.69 | 33 | 16:38 | 20.48 | HAZ | $0.94^{+0.00}_{-0.00}$ | $0.36^{+0.00}_{-0.00}$ | $0.63^{+0.00}_{-0.01}$ | 1.2 | Basalt | 0.015 | 0.087 | 0.267 |
| 1409 | 191.09 | 33 | 17:24 | 9.14 | HAZ | $0.96^{+0.00}_{-0.00}$ | $0.41^{+0.00}_{-0.00}$ | $0.67^{+0.00}_{-0.01}$ | 1.7 | Palagonite | 0.098 | 0.678 | 2.079 |
| 1416 | 195.28 | 33 | 16:38 | 20.72 | HAZ | $0.77^{+0.04}_{-0.05}$ | $-0.02^{+0.07}_{-0.15}$ | $0.76^{+0.06}_{-0.05}$ | 1.6 | Basalt | 0.005 | 0.006 | 2.224 |
| 1418 | 196.50 | 33 | 17:16 | 11.31 | HAZ | $0.95^{+0.01}_{-0.00}$ | $0.41^{+0.00}_{-0.04}$ | $0.59^{+0.03}_{-0.00}$ | 1.7 | Basalt | 0.052 | 0.041 | 3.629 |
| 1422 | 198.93 | 33 | 17:05 | 14.10 | HAZ | $0.94^{+0.01}_{-0.01}$ | $0.10^{+0.04}_{-0.00}$ | $0.86^{+0.00}_{-0.01}$ | 2.3 | Basalt | 0.014 | 0.121 | 1.017 |
| 1444 | 212.57 | 33 | 17:18 | 11.12 | HAZ | $0.94^{+0.02}_{-0.00}$ | $0.06^{+0.04}_{-0.00}$ | $0.84^{+0.03}_{-0.02}$ | 1.8 | Palagonite | 0.007 | 0.005 | 1.672 |
| 1448 | 215.07 | 33 | 16:53 | 17.22 | HAZ | $0.97^{+0.01}_{-0.02}$ | $0.26^{+0.00}_{-0.04}$ | $0.55^{+0.07}_{-0.04}$ | 1.7 | Palagonite | 0.019 | 0.024 | 3.090 |
| 1454 | 218.88 | 33 | 17:04 | 14.60 | HAZ | $0.94^{+0.00}_{-0.00}$ | $0.33^{+0.00}_{-0.00}$ | $0.59^{+0.00}_{-0.00}$ | 1.2 | Basalt | 0.077 | 0.137 | 2.266 |
| 1474 | 231.71 | 33 | 17:21 | 10.57 | HAZ | $0.89^{+0.10}_{-0.01}$ | $0.09^{+0.17}_{-0.04}$ | $0.74^{+0.03}_{-0.24}$ | 0.8 | Palagonite | 0.007 | 0.019 | 0.803 |
| 1480 | 235.57 | 33 | 16:37 | 20.95 | HAZ | $0.89^{+0.04}_{-0.06}$ | $0.38^{+0.04}_{-0.03}$ | $0.52^{+0.08}_{-0.02}$ | 1.7 | Basalt | 0.006 | 0.143 | 2.508 |
| 1484 | 238.19 | 33 | 17:30 | 8.64 | HAZ | $0.97^{+0.02}_{-0.01}$ | $0.30^{+0.01}_{-0.04}$ | $0.71^{+0.05}_{-0.03}$ | 1.7 | Palagonite | 0.008 | 0.367 | 3.090 |
| 1491 | 242.74 | 33 | 17:17 | 11.55 | HAZ | $0.93^{+0.00}_{-0.00}$ | $0.32^{+0.00}_{-0.00}$ | $0.68^{+0.00}_{-0.00}$ | 0.6 | Basalt | 0.075 | 0.020 | 5.356 |
| 1493 | 244.04 | 33 | 17:11 | 13.03 | HAZ | $0.87^{+0.00}_{-0.02}$ | $0.12^{+0.00}_{-0.14}$ | $0.78^{+0.06}_{-0.00}$ | 1.0 | Basalt | 0.196 | 0.167 | 3.394 |
| 1496 | 246.01 | 33 | 17:35 | 7.38 | HAZ | $0.95^{+0.00}_{-0.01}$ | $0.37^{+0.00}_{-0.00}$ | $0.62^{+0.02}_{-0.00}$ | 1.8 | Basalt | 0.031 | 0.645 | 7.379 |
| 1504 | 251.23 | 33 | 17:28 | 8.97 | HAZ | $0.94^{+0.00}_{-0.00}$ | $0.17^{+0.00}_{-0.00}$ | $0.76^{+0.01}_{-0.00}$ | 1.7 | Palagonite | 0.010 | 0.078 | 1.206 |
| 1511 | 255.78 | 33 | 16:42 | 19.61 | HAZ | $0.81^{+0.02}_{-0.03}$ | $0.18^{+0.07}_{-0.13}$ | $0.67^{+0.08}_{-0.05}$ | 1.1 | Basalt | 0.044 | 0.317 | 1.224 |
| 1512 | 256.42 | 33 | 16:28 | 22.63 | HAZ | $0.94^{+0.02}_{-0.37}$ | $0.29^{+0.05}_{-0.89}$ | $0.60^{+0.27}_{-0.05}$ | 1.7 | Basalt | 0.007 | 0.049 | 1.229 |
| 1518 | 260.35 | 33 | 17:02 | 15.07 | HAZ | $0.91^{+0.04}_{-0.02}$ | $0.17^{+0.04}_{-0.04}$ | $0.66^{+0.08}_{-0.11}$ | 0.8 | Palagonite | 0.005 | 0.025 | 3.046 |
| 1537 | 272.68 | 33 | 17:17 | 11.64 | HAZ | $0.92^{+0.00}_{-0.00}$ | $0.28^{+0.00}_{-0.00}$ | $0.50^{+0.00}_{-0.00}$ | 0.7 | Basalt | 0.089 | 0.015 | 2.870 |
| 1555 | 284.18 | 33 | 16:47 | 18.45 | NAV | $0.95^{+0.00}_{-0.00}$ | $0.29^{+0.00}_{-0.00}$ | $0.60^{+0.01}_{-0.00}$ | 1.2 | Basalt | 0.131 | 0.362 | 0.704 |
| 1581 | 300.45 | 33 | 17:14 | 12.33 | HAZ | $0.86^{+0.00}_{-0.01}$ | $0.16^{+0.00}_{-0.00}$ | $0.61^{+0.01}_{-0.00}$ | 2.0 | Basalt | 0.232 | 0.540 | 1.847 |
| 1661 | 346.68 | 33 | 16:44 | 19.20 | HAZ | $0.90^{+0.00}_{-0.01}$ | $0.13^{+0.00}_{-0.07}$ | $0.67^{+0.02}_{-0.01}$ | 1.0 | Basalt | 0.169 | 0.127 | 2.155 |
| 1668 | 350.22 | 33 | 17:15 | 11.29 | HAZ | $0.98^{+0.00}_{-0.02}$ | $0.30^{+0.00}_{-0.05}$ | $0.64^{+0.06}_{-0.00}$ | 1.7 | Palagonite | 0.023 | 0.098 | 1.595 |
| 1675 | 353.89 | 33 | 16:57 | 15.86 | HAZ | $0.94^{+0.01}_{-0.01}$ | $0.36^{+0.04}_{-0.00}$ | $0.63^{+0.01}_{-0.04}$ | 0.6 | Palagonite | 0.020 | 0.016 | 2.803 |
| 1681 | 357.01 | 33 | 16:46 | 18.40 | HAZ | $0.94^{+0.00}_{-0.00}$ | $0.29^{+0.00}_{-0.00}$ | $0.61^{+0.00}_{-0.00}$ | 2.3 | Basalt | 0.109 | 0.012 | 4.206 |
| 1715 | 14.15 | 34 | 16:34 | 20.70 | HAZ | $0.96^{+0.01}_{-0.01}$ | $0.18^{+0.04}_{-0.00}$ | $0.73^{+0.01}_{-0.02}$ | 2.2 | Basalt | 0.021 | 0.159 | 0.787 |
| 1723 | 18.07 | 34 | 17:10 | 11.63 | HAZ | $0.86^{+0.00}_{-0.01}$ | $0.12^{+0.00}_{-0.07}$ | $0.65^{+0.03}_{-0.00}$ | 0.5 | Basalt | 0.228 | 0.432 | 1.943 |
| 1749 | 30.49 | 34 | 16:48 | 16.36 | HAZ | $0.98^{+0.00}_{-0.00}$ | $0.30^{+0.04}_{-0.00}$ | $0.67^{+0.01}_{-0.03}$ | 0.5 | Basalt | 0.389 | 0.706 | 0.791 |

| Sol | Ls[°] | MY | LTST | Sun Elev [°] | Cam | DHG, $g_1$ | DHG, $g_2$ | DHG, α | CYL, D/L | TAB, Dust Analog. | $X^2_{red}$ DHG | $X^2_{red}$ CYL | $X^2_{red}$ TAB |
|---|---|---|---|---|---|---|---|---|---|---|---|---|---|
| 1763 | 37.03 | 34 | 16:16 | 23.67 | HAZ | $0.97^{+0.01}_{-0.02}$ | $0.14^{+0.00}_{-0.04}$ | $0.70^{+0.03}_{-0.01}$ | 2.2 | Basalt | 0.012 | 0.217 | 0.577 |
| 1764 | 37.50 | 34 | 16:46 | 16.40 | HAZ | $0.97^{+0.00}_{-0.00}$ | $0.18^{+0.00}_{-0.00}$ | $0.73^{+0.00}_{-0.00}$ | 2.2 | Basalt | 0.140 | 0.059 | 1.046 |
| 1765 | 37.97 | 34 | 17:16 | 9.12 | HAZ | $0.81^{+0.00}_{-0.01}$ | $-0.08^{+0.00}_{-0.06}$ | $0.82^{+0.02}_{-0.00}$ | 2.2 | Basalt | 0.122 | 0.097 | 2.709 |
| 1770 | 40.27 | 34 | 16:18 | 22.85 | HAZ | $0.98^{+0.00}_{-0.02}$ | $0.14^{+0.00}_{-0.04}$ | $0.73^{+0.02}_{-0.01}$ | 2.2 | Basalt | 0.017 | 0.220 | 0.500 |
| 1771 | 40.74 | 34 | 16:48 | 15.68 | HAZ | $0.97^{+0.00}_{-0.00}$ | $0.18^{+0.00}_{-0.00}$ | $0.78^{+0.00}_{-0.00}$ | 0.5 | Basalt | 0.107 | 0.335 | 0.706 |
| 1772 | 41.21 | 34 | 17:19 | 8.44 | HAZ | $0.84^{+0.00}_{-0.01}$ | $-0.09^{+0.00}_{-0.10}$ | $0.84^{+0.02}_{-0.00}$ | 0.8 | Basalt | 0.086 | 0.008 | 3.742 |
| 1777 | 43.49 | 34 | 16:20 | 22.06 | HAZ | $0.95^{+0.03}_{-0.01}$ | $0.10^{+0.04}_{-0.04}$ | $0.75^{+0.02}_{-0.03}$ | 2.2 | Basalt | 0.014 | 0.136 | 0.458 |
| 1779 | 44.42 | 34 | 17:21 | 7.71 | HAZ | $0.84^{+0.01}_{-0.01}$ | $0.05^{+0.10}_{-0.03}$ | $0.81^{+0.02}_{-0.04}$ | 1.1 | Basalt | 0.097 | 0.032 | 3.325 |
| 1791 | 49.88 | 34 | 16:31 | 19.26 | HAZ | $0.93^{+0.05}_{-0.03}$ | $0.09^{+0.09}_{-0.08}$ | $0.79^{+0.04}_{-0.07}$ | 2.2 | Basalt | 0.008 | 0.023 | 0.321 |
| 1802 | 54.86 | 34 | 16:06 | 24.52 | HAZ | $0.89^{+0.00}_{-0.01}$ | $0.38^{+0.04}_{-0.04}$ | $0.70^{+0.03}_{-0.02}$ | 1.1 | Basalt | 0.031 | 0.174 | 0.525 |
| 1805 | 56.23 | 34 | 16:38 | 17.14 | HAZ | $0.97^{+0.00}_{-0.00}$ | $0.49^{+0.00}_{-0.00}$ | $0.66^{+0.01}_{-0.00}$ | 1.9 | Basalt | 0.156 | 0.211 | 0.281 |
| 1816 | 61.19 | 34 | 17:16 | 8.31 | HAZ | $0.82^{+0.01}_{-0.01}$ | $0.15^{+0.07}_{-0.03}$ | $0.74^{+0.03}_{-0.04}$ | 1.0 | Palagonite | 0.014 | 0.020 | 1.437 |
| 1818 | 62.08 | 34 | 16:49 | 14.39 | HAZ | $0.94^{+0.00}_{-0.01}$ | $0.13^{+0.00}_{-0.04}$ | $0.78^{+0.02}_{-0.01}$ | 2.2 | Basalt | 0.025 | 0.110 | 0.771 |
| 1821 | 63.42 | 34 | 16:28 | 19.01 | HAZ | $0.96^{+0.00}_{-0.00}$ | $0.33^{+0.00}_{-0.00}$ | $0.60^{+0.00}_{-0.00}$ | 1.2 | Basalt | 0.102 | 0.510 | 0.585 |
| 1824 | 64.78 | 34 | 16:55 | 12.89 | HAZ | $0.92^{+0.00}_{-0.01}$ | $0.09^{+0.00}_{-0.04}$ | $0.80^{+0.02}_{-0.00}$ | 2.2 | Basalt | 0.035 | 0.054 | 0.889 |
| 1831 | 67.92 | 34 | 17:07 | 10.12 | HAZ | $0.76^{+0.01}_{-0.02}$ | $-0.36^{+0.06}_{-0.08}$ | $0.85^{+0.01}_{-0.01}$ | 1.4 | Basalt | 0.012 | 0.031 | 2.634 |
| 1836 | 70.17 | 34 | 17:18 | 7.53 | HAZ | $0.86^{+0.01}_{-0.00}$ | $0.09^{+0.04}_{-0.00}$ | $0.77^{+0.00}_{-0.02}$ | 1.8 | Basalt | 0.147 | 0.016 | 2.852 |
| 1838 | 71.06 | 34 | 17:01 | 11.49 | HAZ | $0.72^{+0.00}_{-0.00}$ | $-0.31^{+0.03}_{-0.00}$ | $0.86^{+0.00}_{-0.01}$ | 0.5 | Basalt | 0.064 | 0.398 | 1.282 |
| 1839 | 71.49 | 34 | 16:09 | 22.90 | HAZ | $0.94^{+0.04}_{-0.01}$ | $0.02^{+0.04}_{-0.00}$ | $0.72^{+0.01}_{-0.04}$ | 2.3 | Basalt | 0.011 | 0.012 | 0.822 |
| 1845 | 74.19 | 34 | 16:41 | 15.71 | HAZ | $0.76^{+0.00}_{-0.01}$ | $-0.48^{+0.03}_{-0.12}$ | $0.91^{+0.011}_{-0.00}$ | 2.0 | Basalt | 0.087 | 0.160 | 0.970 |
| 1848 | 75.53 | 34 | 16:19 | 20.55 | HAZ | $0.95^{+0.00}_{-0.00}$ | $0.14^{+0.00}_{-0.00}$ | $0.74^{+0.00}_{-0.00}$ | 2.0 | Basalt | 0.021 | 0.295 | 0.782 |
| 1849 | 75.81 | 34 | 07:27 | 17.74 | NAV | $0.86^{+0.00}_{-0.00}$ | $-0.05^{+0.04}_{-0.04}$ | $0.85^{+0.01}_{-0.01}$ | 2.3 | Basalt | 0.050 | 0.110 | 0.231 |
| 1853 | 77.79 | 34 | 16:54 | 12.74 | HAZ | $0.70^{+0.01}_{-0.00}$ | $-0.33^{+0.05}_{-0.06}$ | $0.88^{+0.01}_{-0.01}$ | 1.0 | Basalt | 0.077 | 0.231 | 1.010 |
| 1859 | 80.47 | 34 | 16:19 | 20.44 | HAZ | $0.97^{+0.00}_{-0.01}$ | $0.14^{+0.00}_{-0.04}$ | $0.69^{+0.01}_{-0.00}$ | 2.0 | Basalt | 0.203 | 0.141 | 0.560 |
| 1863 | 82.28 | 34 | 16:48 | 14.11 | HAZ | $0.71^{+0.01}_{-0.00}$ | $-0.33^{+0.05}_{-0.09}$ | $0.89^{+0.01}_{-0.01}$ | 2.0 | Basalt | 0.114 | 0.360 | 0.888 |
| 1865 | 83.18 | 34 | 16:42 | 15.46 | HAZ | $0.81^{+0.00}_{-0.00}$ | $0.38^{+0.00}_{-0.00}$ | $0.64^{+0.00}_{-0.00}$ | 2.1 | Basalt | 0.257 | 0.060 | 0.493 |
| 1872 | 86.34 | 34 | 17:01 | 11.25 | NAV | $0.85^{+0.01}_{-0.01}$ | $-0.23^{+0.07}_{-0.00}$ | $0.91^{+0.00}_{-0.01}$ | 2.3 | Basalt | 0.029 | 0.037 | 0.372 |
| 1879 | 89.51 | 34 | 17:02 | 10.81 | HAZ | $0.97^{+0.00}_{-0.00}$ | $0.06^{+0.00}_{-0.00}$ | $0.88^{+0.00}_{-0.00}$ | 2.0 | Basalt | 0.027 | 0.101 | 0.732 |
| 1885 | 92.21 | 34 | 16:05 | 23.33 | HAZ | $0.92^{+0.00}_{-0.01}$ | $0.13^{+0.00}_{-0.04}$ | $0.70^{+0.02}_{-0.01}$ | 2.3 | Basalt | 0.031 | 0.047 | 0.639 |
| 1886 | 92.68 | 34 | 16:59 | 11.61 | HAZ | $0.66^{+0.04}_{-0.03}$ | $-0.15^{+0.19}_{-0.17}$ | $0.88^{+0.04}_{-0.06}$ | 2.2 | Basalt | 0.021 | 0.095 | 1.239 |
| 1892 | 95.41 | 34 | 16:37 | 16.48 | HAZ | $0.97^{+0.00}_{-0.00}$ | $0.26^{+0.00}_{-0.00}$ | $0.71^{+0.00}_{-0.00}$ | 2.0 | Basalt | 0.069 | 0.349 | 0.497 |
| 1894 | 96.14 | 34 | 07:12 | 14.36 | NAV | $0.90^{+0.00}_{-0.00}$ | $0.02^{+0.00}_{-0.04}$ | $0.85^{+0.01}_{-0.00}$ | 2.3 | Basalt | 0.065 | 0.023 | 0.166 |
| 1895 | 96.78 | 34 | 16:37 | 16.40 | HAZ | $0.78^{+0.01}_{-0.00}$ | $-0.40^{+0.06}_{-0.00}$ | $0.91^{+0.00}_{-0.01}$ | 2.0 | Basalt | 0.101 | 0.038 | 0.874 |
| 1902 | 100.00 | 34 | 17:12 | 8.73 | HAZ | $0.89^{+0.08}_{-0.01}$ | $0.24^{+0.08}_{-0.10}$ | $0.66^{+0.20}_{-0.16}$ | 0.8 | Palagonite | 0.006 | 0.017 | 0.876 |
| 1904 | 100.92 | 34 | 17:04 | 10.62 | HAZ | $0.85^{+0.01}_{-0.02}$ | $0.16^{+0.00}_{-0.04}$ | $0.62^{+0.03}_{-0.01}$ | 2.3 | Basalt | 0.063 | 0.037 | 1.854 |
| 1911 | 104.16 | 34 | 17:16 | 7.90 | HAZ | $0.85^{+0.00}_{-0.00}$ | $-0.05^{+0.03}_{-0.00}$ | $0.82^{+0.00}_{-0.01}$ | 0.7 | Basalt | 0.080 | 0.040 | 2.814 |
| 1916 | 106.47 | 34 | 16:27 | 19.00 | NAV | $0.85^{+0.01}_{-0.01}$ | $0.02^{+0.07}_{-0.10}$ | $0.83^{+0.03}_{-0.03}$ | 1.0 | Basalt | 0.086 | 0.036 | 0.237 |
| 1924 | 110.02 | 34 | 06:59 | 11.59 | NAV | $0.90^{+0.01}_{-0.00}$ | $0.09^{+0.04}_{-0.00}$ | $0.82^{+0.00}_{-0.02}$ | 2.3 | Basalt | 0.034 | 0.152 | 0.439 |
| 1925 | 110.69 | 34 | 17:06 | 10.39 | HAZ | $0.90^{+0.01}_{-0.03}$ | $0.35^{+0.00}_{-0.05}$ | $0.50^{+0.04}_{-0.00}$ | 0.8 | Basalt | 0.296 | 0.038 | 1.505 |
| 1927 | 111.62 | 34 | 16:53 | 13.18 | HAZ | $0.91^{+0.06}_{-0.04}$ | $-0.13^{+0.07}_{-0.07}$ | $0.87^{+0.01}_{-0.02}$ | 2.0 | Basalt | 0.011 | 0.066 | 1.157 |
| 1928 | 112.08 | 34 | 16:20 | 20.77 | HAZ | $0.95^{+0.03}_{-0.07}$ | $0.06^{+0.04}_{-0.08}$ | $0.76^{+0.07}_{-0.04}$ | 2.2 | Basalt | 0.006 | 0.040 | 0.286 |
| 1929 | 112.58 | 34 | 17:31 | 4.75 | HAZ | $0.95^{+0.00}_{-0.00}$ | $0.37^{+0.00}_{-0.00}$ | $0.68^{+0.00}_{-0.01}$ | 1.7 | Palagonite | 0.084 | 0.673 | 2.514 |
| 1932 | 113.98 | 34 | 16:50 | 14.01 | HAZ | $0.93^{+0.00}_{-0.00}$ | $0.13^{+0.00}_{-0.00}$ | $0.76^{+0.00}_{-0.00}$ | 2.0 | Basalt | 0.202 | 0.114 | 0.637 |
| 1934 | 114.94 | 34 | 17:23 | 6.51 | HAZ | $0.90^{+0.00}_{-0.00}$ | $0.17^{+0.04}_{-0.00}$ | $0.76^{+0.00}_{-0.02}$ | 1.8 | Palagonite | 0.083 | 0.057 | 1.790 |
| 1935 | 115.41 | 34 | 16:53 | 13.45 | HAZ | $0.66^{+0.01}_{-0.01}$ | $-0.66^{+0.01}_{-0.01}$ | $0.90^{+0.00}_{-0.01}$ | 2.0 | Basalt | 0.135 | 0.071 | 0.722 |
| 1937 | 116.36 | 34 | 16:53 | 13.51 | HAZ | $0.94^{+0.00}_{-0.00}$ | $0.17^{+0.00}_{-0.00}$ | $0.77^{+0.01}_{-0.00}$ | 2.0 | Basalt | 0.141 | 0.057 | 0.427 |
| 1938 | 116.83 | 34 | 16:57 | 12.56 | HAZ | $0.70^{+0.00}_{-0.01}$ | $-0.70^{+0.03}_{-0.00}$ | $0.86^{+0.00}_{-0.01}$ | 2.0 | Basalt | 0.011 | 0.077 | 0.495 |
| 1947 | 121.15 | 34 | 17:04 | 11.12 | HAZ | $0.63^{+0.00}_{-0.00}$ | $-0.63^{+0.00}_{-0.00}$ | $0.95^{+0.00}_{-0.00}$ | 2.1 | Basalt | 0.268 | 0.401 | 0.275 |
| 1963 | 128.95 | 34 | 17:27 | 6.18 | HAZ | $0.91^{+0.00}_{-0.00}$ | $0.17^{+0.00}_{-0.00}$ | $0.85^{+0.00}_{-0.00}$ | 0.6 | Palagonite | 0.130 | 0.392 | 0.329 |
| 1964 | 129.44 | 34 | 17:18 | 8.25 | HAZ | $0.93^{+0.01}_{-0.00}$ | $0.32^{+0.00}_{-0.00}$ | $0.76^{+0.01}_{-0.01}$ | 0.6 | Palagonite | 0.077 | 0.123 | 0.750 |
| 1968 | 131.40 | 34 | 16:23 | 21.03 | NAV | $0.78^{+0.01}_{-0.01}$ | $-0.18^{+0.09}_{-0.06}$ | $0.89^{+0.01}_{-0.02}$ | 0.8 | Basalt | 0.011 | 0.013 | 0.251 |
| 1969 | 131.91 | 34 | 17:05 | 11.42 | HAZ | $0.72^{+0.00}_{-0.01}$ | $-0.72^{+0.01}_{-0.00}$ | $0.87^{+0.01}_{-0.00}$ | 2.0 | Basalt | 0.100 | 0.161 | 0.246 |
| 1971 | 132.70 | 34 | 07:12 | 15.48 | NAV | $0.89^{+0.01}_{-0.00}$ | $-0.05^{+0.04}_{-0.04}$ | $0.88^{+0.01}_{-0.01}$ | 1.9 | Palagonite | 0.147 | 0.081 | 1.123 |
| 1971 | 132.91 | 34 | 17:14 | 9.28 | HAZ | $0.76^{+0.01}_{-0.02}$ | $-0.11^{+0.03}_{-0.15}$ | $0.88^{+0.03}_{-0.01}$ | 1.1 | Basalt | 0.038 | 0.043 | 0.197 |
| 1972 | 133.40 | 34 | 16:45 | 16.04 | HAZ | $0.95^{+0.00}_{-0.00}$ | $0.25^{+0.00}_{-0.00}$ | $0.76^{+0.00}_{-0.00}$ | 1.9 | Basalt | 0.203 | 0.190 | 0.154 |
| 1973 | 133.91 | 34 | 17:03 | 11.86 | HAZ | $0.97^{+0.00}_{-0.00}$ | $0.30^{+0.00}_{-0.00}$ | $0.78^{+0.00}_{-0.00}$ | 2.2 | Basalt | 0.231 | 0.178 | 0.266 |
| 1974 | 134.41 | 34 | 17:12 | 9.90 | HAZ | $0.74^{+0.01}_{-0.00}$ | $-0.37^{+0.09}_{-0.03}$ | $0.93^{+0.00}_{-0.01}$ | 0.5 | Basalt | 0.007 | 0.067 | 0.982 |
| 1975 | 134.90 | 34 | 16:39 | 17.55 | HAZ | $0.96^{+0.00}_{-0.00}$ | $0.33^{+0.00}_{-0.00}$ | $0.70^{+0.00}_{-0.00}$ | 1.9 | Basalt | 0.292 | 0.434 | 0.228 |
| 1978 | 136.41 | 34 | 16:49 | 15.28 | HAZ | $0.98^{+0.00}_{-0.00}$ | $0.38^{+0.00}_{-0.00}$ | $0.73^{+0.00}_{-0.00}$ | 0.5 | Basalt | 0.140 | 0.319 | 0.166 |
| 1979 | 136.93 | 34 | 17:25 | 6.87 | HAZ | $0.96^{+0.00}_{-0.00}$ | $0.33^{+0.00}_{-0.00}$ | $0.78^{+0.00}_{-0.00}$ | 1.7 | Palagonite | 0.154 | 0.775 | 0.911 |
| 1984 | 139.46 | 34 | 17:10 | 10.45 | HAZ | $0.73^{+0.01}_{-0.01}$ | $-0.31^{+0.09}_{-0.08}$ | $0.92^{+0.01}_{-0.01}$ | 0.5 | Basalt | 0.010 | 0.118 | 1.085 |
| 1988 | 141.50 | 34 | 17:29 | 6.15 | HAZ | $0.90^{+0.00}_{-0.00}$ | $-0.02^{+0.04}_{-0.00}$ | $0.84^{+0.00}_{-0.01}$ | 1.8 | Palagonite | 0.244 | 0.615 | 1.789 |
| 1989 | 142.00 | 34 | 16:30 | 20.29 | HAZ | $0.93^{+0.02}_{-0.05}$ | $-0.13^{+0.04}_{-0.14}$ | $0.85^{+0.04}_{-0.02}$ | 2.1 | Basalt | 0.005 | 0.165 | 0.429 |
| 1998 | 146.66 | 34 | 17:03 | 12.47 | NAV | $0.92^{+0.00}_{-0.00}$ | $-0.17^{+0.00}_{-0.00}$ | $0.94^{+0.00}_{-0.00}$ | 0.5 | Palagonite | 0.032 | 0.020 | 0.442 |
| 2000 | 147.69 | 34 | 16:49 | 16.02 | HAZ | $0.97^{+0.00}_{-0.07}$ | $-0.06^{+0.00}_{-0.14}$ | $0.86^{+0.04}_{-0.00}$ | 2.0 | Basalt | 0.006 | 0.071 | 0.927 |
| 2001 | 148.22 | 34 | 17:05 | 12.04 | HAZ | $0.89^{+0.03}_{-0.03}$ | $-0.27^{+0.07}_{-0.06}$ | $0.93^{+0.00}_{-0.00}$ | 2.0 | Palagonite | 0.006 | 0.025 | 1.197 |